\documentclass[aps,prd,reprint,nofootinbib,eqsecnum,showpacs]{revtex4-1}
\let\revappendix\appendix                                 % Make revtex's \appendix* compatible with package cleveref (see below)
\usepackage{amssymb,amsmath,amsfonts}                     % American Mathematical Society packages for symbols, math, and fonts
\usepackage{mathtools}                                    % Extension to amsmath
\usepackage[mathscr]{euscript}                            % Euler Script font: \mathscr{}
\usepackage{bbm}                                          % \mathbbm{}, \mathbbmss{}, \mathbbmtt{}
\usepackage{bm}                                           % Bold symbols: \bm{}
\usepackage{graphicx}                                     % Extended Graphics
\graphicspath{{figures/}}                                 % Graphics path
\usepackage{hyperref}                                     % Hypertext links -- when used with cleveref, it must be loaded first
\usepackage[capitalise]{cleveref}                         % Clever reference handling -- use with \cref{}
\usepackage{rotating}                                     % Float rotation
\usepackage{xstring}                                      % String manipulation
\usepackage{color}                                        % Color
\definecolor{DarkBlue}{rgb}{0,0.1,0.7}                    % Define link color (color or xcolor package required)
\hypersetup{
	bookmarksnumbered=true,
	citecolor=DarkBlue,
	colorlinks=true,
	linkcolor=DarkBlue,
	urlcolor=DarkBlue
}
\newcommand{\crefrangeconjunction}{--}                    % Put a dash instead of 'to' when linking equations in \cref{eq1,eq2,...}
\newcommand{\creflastconjunction}{, and\nobreakspace}     % Use of the serial comma for multiple references
\crefname{section}{Sec.}{Secs.}                           % section reference (mid-sentence)
\Crefname{section}{Section}{Sections}                     % section reference (sentence beginning)
\crefformat{subequations}{Eqs.~(#2#1#3)}                  % subequations reference format (mid-sentence)
\Crefformat{subequations}{Equations~(#2#1#3)}             % subequations reference format (sentence begining)
\crefformat{appsec}{the #2Appendix#3}                     % single appendix reference (mid-sentence) -- use with \label[appsec]{}
\Crefformat{appsec}{#2Appendix#3}                         % single appendix reference (sentence begining) --use with \label[appsec]{}
\newcommand{\refcite}[1]{%                                % Put Ref(s). before citation(s) (mid-sentence)
	\begingroup
		\def\tempx{0}%
		\StrCount{#1}{,}[\tempx]%
		\ifnum\tempx > 0
			Refs.~\cite{#1}%
		\else
			Ref.~\cite{#1}%
		\fi
	\endgroup
}
\newcommand{\Refcite}[1]{%                                % Put Reference(s) before citation(s) (sentence beginning)
	\begingroup
		\def\tempx{0}%
		\StrCount{#1}{,}[\tempx]%
		\ifnum\tempx > 0 
			References~\cite{#1}%
		\else
			Reference~\cite{#1}%
		\fi
	\endgroup
}
\newcommand{\der}[2]{\frac{\mathrm{d}#1}{\mathrm{d}#2}}   % Ordinary derivative

\begin{document}
%
%-----------------------Title-----------------------------%
%
\title{Saturation of the \texorpdfstring{$f$}{f}-mode instability in neutron stars\texorpdfstring{\\}{:} II. Applications and results}
\author{Pantelis Pnigouras}
	%\email{}
	\affiliation{Theoretical Astrophysics, IAAT, Eberhard-Karls University of T\"ubingen, 72076 T\"ubingen, Germany}
\author{Kostas D. Kokkotas}
	\affiliation{Theoretical Astrophysics, IAAT, Eberhard-Karls University of T\"ubingen, 72076 T\"ubingen, Germany}
\date{\today}
%
%----------------------Abstract---------------------------%

\begin{abstract}
	We present the first results on the saturation of the $f$-mode instability in neutron stars, due to nonlinear mode coupling. Emission of gravitational waves drives the $f$-mode (fundamental mode) unstable in fast-rotating, newborn neutron stars. The initial growth phase of the mode is followed by its saturation, because of energy leaking to other modes of the star. The saturation point determines the strain of the generated gravitational-wave signal, which can then be used to extract information about the neutron star equation of state. The parent (unstable) mode couples via parametric resonances with pairs of daughter modes, with the triplets' evolution exhibiting a rich variety of behaviors. We study both supernova- and merger-derived neutron stars, simply modeled as polytropes in a Newtonian context, and show that the parent may couple to many different daughter pairs during the star's evolution through the instability window, with the saturation amplitude changing by orders of magnitude.
\end{abstract}

%--------------------PACS numbers-------------------------%

\pacs{04.30.Db, 04.40.Dg, 97.10.Sj, 97.60.Jd}

\maketitle

%-----------------------------------------------------------------------------------------------------------------------------%
%%%%%%%%%%%%%%%%%%%%%%%%%%%%%%%%%%%%%%%%%%%%%%%%%%%%%%%%%%%%%%%%%%%%%%%%%%%%%%%%%%%%%%%%%%%%%%%%%%%%%%%%%%%%%%%%%%%%%%%%%%%%%%%
%-----------------------------------------------------------------------------------------------------------------------------%

\section{Introduction} \label{sec:Introduction}

It has been known since the 1970s that nonradial, stellar oscillation modes can be driven unstable due to the emission of gravitational waves, thanks to the Chandrasekhar-Friedman-Schutz (CFS) mechanism \cite{Chandrasekhar1970,FriedmanSchutz1978,*FriedmanSchutz1978b}. In rapidly rotating neutron stars, low multipoles could be driven unstable, producing a significant amount of gravitational radiation \cite{IpserLindblom1990,*IpserLindblom1991}. Now that the gravitational wave window to space has been opened \cite{AbbottEtAl2016}, these signals may provide useful information about the neutron star interior \cite{AnderssonKokkotas1996,AnderssonKokkotas1998,KokkotasEtAl2001,BenharEtAl2004,GaertigKokkotas2011,DonevaEtAl2013,DonevaKokkotas2015}; gravitational wave asteroseismology is expected to give some answers about the equation of state of matter at supranuclear densities, avoiding the problems that QCD and terrestrial experiments are currently facing.

Nevertheless, there is still uncertainty behind the process saturating the instability. Studies on the CFS-unstable $r$-modes (horizontal fluid motions driven by rotation \cite{PapaloizouPringle1978}) showed that nonlinear mode coupling saturates the instability quite efficiently \cite{SchenkEtAl2001,Morsink2002,ArrasEtAl2003,BrinkEtAl2004,*BrinkEtAl2004b,*BrinkEtAl2005}. Other proposed saturation mechanisms, like large-amplitude viscous effects \cite{AlfordEtAl2012,PassamontiGlampedakis2012} or the interaction of superfluid vortices with superconducting flux tubes \cite{HaskellEtAl2014}, may lead to higher or lower saturation amplitudes, respectively.

We focus on another class of potentially unstable modes, the $f$-modes (fundamental oscillations), which comprise large-scale density perturbations. Although the $f$-mode instability is active in a much smaller part of the parameter space (instability window) and has a longer growth time, compared to the $r$-mode, the calculation of its saturation amplitude is still important for the evolution of newborn neutron stars \cite{PassamontiEtAl2013}. Recent studies \cite{DonevaEtAl2015} have also shown a very promising scenario, where an $f$-mode instability could develop very fast in post-merger neutron star remnants.

So far, no robust estimate has been provided for the saturation amplitude of unstable $f$-modes. Some upper limits have been derived by \refcite{KastaunEtAl2010}, where the effects of nonlinear damping (like wave breaking) were mainly studied, using a general relativistic simulation under the Cowling approximation (fixed spacetime approximation; for a similar study on $r$-modes, see also \refcite{Kastaun2011}). In the aforementioned study, mode coupling was also observed, for the quadrupole $f$-mode ($l=m=2$, where $l$ and $m$ are the degree and order, respectively, of the spherical harmonic $Y_l^m$ that describes the mode).

However, the CFS instability sets in on secular time scales, way beyond the current capabilities of nonlinear hydrodynamic simulations. Given the large time needed for the unstable mode to grow, simulations resort to starting with high mode-amplitude values and then tracking the amplitude decay. This process gives an upper bound for the saturation amplitude, which could still be far, though, from the actual value. As shown by \refcite{SchenkEtAl2001,Morsink2002,ArrasEtAl2003,BrinkEtAl2004,*BrinkEtAl2004b,*BrinkEtAl2005}, nonlinear mode coupling ought to operate at quite low amplitudes, rendering other effects, like the wave breaking of \refcite{KastaunEtAl2010} and the large-amplitude viscous dissipation of \refcite{AlfordEtAl2012,PassamontiGlampedakis2012}, unimportant. On the other hand, the saturation mechanism described in \refcite{HaskellEtAl2014} may lead to lower saturation values, but is relevant only for mature stars.

In this paper, we present results on the $f$-mode saturation, using quadratic mode coupling to other polar modes, for Newtonian polytropic stars. The formalism and its main implications were described in detail in a previous paper \cite{PnigourasKokkotas2015} (henceforth Paper I).

As opposed to the linear approximation, which gives rise to the oscillation spectrum of the star (e.g., \refcite{UnnoEtAl1989,AertsEtAl2010}), quadratic perturbations build up a three-mode-interaction network, in which the modes of the star couple in triplets. The (complex) mode amplitudes $Q_i$ of a given triplet are then described by
\begin{subequations}
	\label[subequations]{equations of motion with normalisation choice}
	\begin{align}
		\dot{Q}_\alpha &= \gamma_\alpha Q_\alpha+\frac{i\omega_\alpha\mathcal{H}}{E_\mathrm{unit}}Q_\beta Q_\gamma e^{-i\Delta\omega t}, \label{mode alpha equation of motion with normalisation choice} \\
		\dot{Q}_\beta &= \gamma_\beta Q_\beta+\frac{i\omega_\beta\mathcal{H}}{E_\mathrm{unit}} Q_\gamma^* Q_\alpha e^{i\Delta\omega t}, \label{mode beta equation of motion with normalisation choice} \\
		\dot{Q}_\gamma &= \gamma_\gamma Q_\gamma+\frac{i\omega_\gamma\mathcal{H}}{E_\mathrm{unit}} Q_\alpha Q_\beta^* e^{i\Delta\omega t}, \label{mode gamma equation of motion with normalisation choice}
	\end{align}
\end{subequations}
where the parameters $\gamma_i$ are the linear growth/damping rates, $\omega_i$ the mode frequencies, $\mathcal{H}$ the coupling coefficient, $E_\mathrm{unit}$ the mode energy at unit amplitude (based on the normalization choice), and $\Delta\omega=\omega_\alpha-\omega_\beta-\omega_\gamma$ the detuning parameter.

The coupling occurs when i) an internal resonance exists between the three modes, of the form $\omega_\alpha\approx\omega_\beta+\omega_\gamma$, and ii) a set of selection rules is satisfied for the degrees $l_i$ and the orders $m_i$ of the modes \cite{Dziembowski1982,PnigourasKokkotas2015}. The first condition guarantees that the oscillatory dependence of the nonlinear term is very slow, so that it contributes to the long-term dynamics of the system, whereas the second condition is what makes the coupling coefficient $\mathcal{H}$ nonzero (see Paper I and references therein).

Coupling of an unstable $(\gamma_\alpha>0)$ mode to two other, stable $(\gamma_{\beta,\gamma}<0)$, modes can lead to a \emph{parametric resonance instability}: the unstable (parent) mode grows until its amplitude surpasses the \emph{parametric instability threshold} (PIT). At this point, the stable (daughter) modes, coupled to the parent, start growing by draining energy from it. Eventually, and if certain stability conditions are met, the system will reach an equilibrium and saturate (\cref{fig:parametrically resonant system}\hyperref[fig:parametrically resonant system]{a}). The parent mode saturates close to the PIT, which is given by \cite{PnigourasKokkotas2015,Dziembowski1982}
\begin{equation}
	|Q_\mathrm{PIT}|^2=\frac{\gamma_\beta \gamma_\gamma}{\omega_\beta \omega_\gamma}\frac{E_\mathrm{unit}^2}{\mathcal{H}^2}\left[1+\left(\frac{\Delta\omega}{\gamma_\beta+\gamma_\gamma}\right)^2\right]. \label{PIT}
\end{equation}

As described before, a tacit consequence of quadratic nonlinearities is that modes couple in triplets. This means that individual couplings consist of three modes only, with the daughter modes trying to stop the growth of the parent mode. Of course, the same parent can couple to more than one pairs of daughters. However, not all couplings become important. Remember that, until the PIT is crossed, the parent does not really ``feel'' the presence of the daughters. Since each coupled triplet has its own PIT, only the couplings with the lowest PITs will affect the parent's evolution. In fact, as we shall see later on, the triplet with the lowest PIT is usually the one that determines the parent's saturation amplitude.

Following this paradigm, we find the couplings of an unstable $f$-mode to other polar modes and then calculate its saturation amplitude, throughout the instability window. The setup of this process is presented in \cref{sec:Setup}. The results for both supernova-derived neutron stars and post-merger remnants can be found in \cref{sec:Results}. In \cref{sec:The saturation mechanism} we review the details behind the saturation mechanism of the parametrically resonant system \eqref{equations of motion with normalisation choice}. Comparison with previous work on the saturation of the $r$-mode instability via mode coupling is discussed in \cref{sec:Comparison with r-modes}. We conclude with a summary and some final remarks in \cref{sec:Summary}.

%-----------------------------------------------------------------------------------------------------------------------------%
%%%%%%%%%%%%%%%%%%%%%%%%%%%%%%%%%%%%%%%%%%%%%%%%%%%%%%%%%%%%%%%%%%%%%%%%%%%%%%%%%%%%%%%%%%%%%%%%%%%%%%%%%%%%%%%%%%%%%%%%%%%%%%%
%-----------------------------------------------------------------------------------------------------------------------------%

\section{Setup} \label{sec:Setup}

We can obtain the saturation amplitude of an unstable $f$-mode, for a specific temperature $T$ and angular velocity $\Omega$ of the star, by following these steps:
\begin{enumerate}
	\setlength{\itemsep}{0pt}
	\item[A.] Calculate mode eigenfrequencies $\omega$ and eigenfunctions $\bm{\xi}$.
	\item[B.] Calculate mode growth/damping rates $\gamma$.
	\item[C.] Find all possible mode couplings and calculate their PITs.
	\item[D.] Locate the triplet with the lowest PIT and check whether saturation is successful.
\end{enumerate}
Repeating this for a grid of $(T,\Omega)$ pairs, we get the unstable mode's saturation amplitude throughout the instability window. Below in this section we are going to review every step in more detail.

%-----------------------------------------------------------------------------------------------------------------------------%
%-----------------------------------------------------------------------------------------------------------------------------%

\subsection{Eigenfrequencies and eigenfunctions} \label{subsec:Eigenfrequencies and eigenfunctions}

As described in Paper I, we use the slow-rotation approximation in order to determine the mode frequencies. The main reason for this is that we want as many modes as possible to be available for coupling, and solving for the exact eigenfrequencies and eigenfunctions of a rotating star can be quite cumbersome, if one wants to obtain many modes. Details about the validity of this approximation are discussed below.

All polar modes with degrees $l\le 11$ and overtones $n\le 10$ are acquired, in the nonrotating limit, by use of a shooting-to-a-fitting-point method \cite{PressEtAl1992,*PressEtAl1996}. Higher overtones and multipoles were harder to obtain, due to numerical issues. Whether these modes are enough is going to be addressed in retrospect, in \cref{sec:Results}.

The eigenfrequencies are then corrected due to rotation as \cite{Saio1981}
\begin{equation}
	\omega=\omega_0+mC_1\Omega+\frac{C_2}{\omega_0}\Omega^2+\mathcal{O}(\Omega^3), \label{rotationally corrected eigenfrequency}
\end{equation} 
where $\omega_0$ is the frequency at the nonrotating limit, and $C_1$, $C_2$ parameters that depend on the equation of state and mode properties. Although we do calculate the rotationally corrected (to first order) eigenfunctions $[\bm{\xi}=\bm{\xi}_0+\bm{\xi}_1+\bm{\mathcal{O}}(\Omega^2)]$ \cite{Saio1981}, we are using the nonrotating solutions $\bm{\xi}_0$ to obtain the various mode quantities, like growth/damping rates and coupling coefficients (more details about this are discussed later on in this section).

\begin{figure*}
	\includegraphics[width=\textwidth,keepaspectratio]{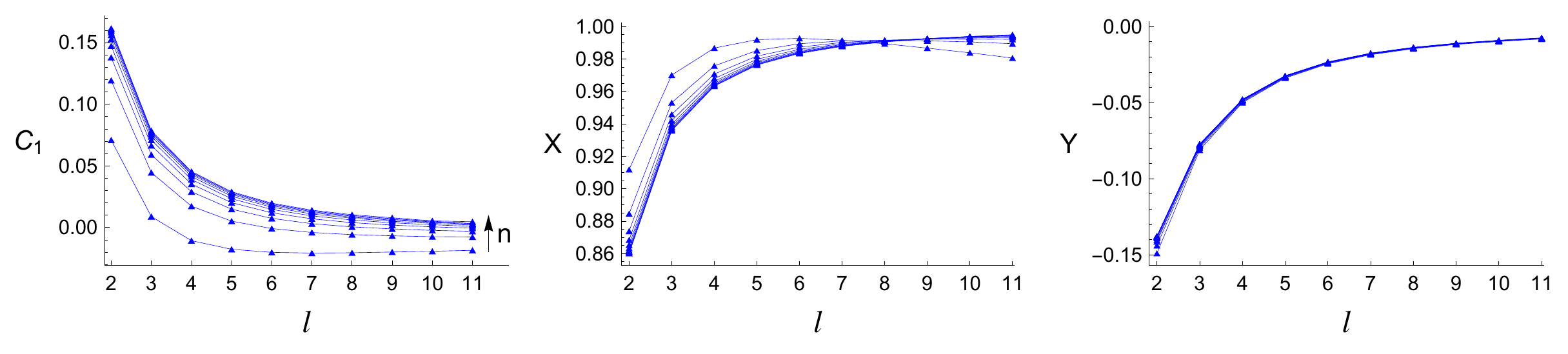}
	\caption{Rotational correction parameters $C_1$ and $C_2=X+m^2 Y$ for $g$-modes with $n\le 10$, as functions of the degree $l$. For increasing $n$, $C_1\rightarrow [l(l+1)]^{-1}$. For increasing $l$ and $n$, $X\rightarrow 1$ and $Y\rightarrow 0$. A polytrope with a polytropic exponent $\Gamma=2$ and an adiabatic exponent $\Gamma_1=2.1$ was used, although this behavior does not seem to change for different values.}
	\label{fig:model_7_gmode_eigenfrequency_corrections}
\end{figure*}

Higher than second-order corrections to the eigenfrequencies should become important at large angular velocities of the star. In fact, $\mathcal{O}(\Omega^2)$ corrections for $g$-modes (buoyancy modes) are divergent as the overtone increases (i.e., as $\omega_0\rightarrow 0$). The parameter $C_2$ can be decomposed as $X+m^2 Y$ \cite{Saio1981}. For $g$-modes with increasing $l$ and $n$, $X\rightarrow 1$ and $Y\rightarrow 0$ (see \cref{fig:model_7_gmode_eigenfrequency_corrections}), so, in this case, second-order corrections scale as $\omega_0^{-1}$. This behavior seems to be independent from the (polytropic) equation of state in use.

The validity of the slow-rotation approximation for $g$-modes can be seen in much detail in \refcite{BallotEtAl2010}, where simulations were performed for a $\Gamma=4/3$ polytrope and rotational corrections up to third order were calculated.\footnote{Based on the results given there, we cannot verify the behavior of \cref{fig:model_7_gmode_eigenfrequency_corrections} for the $\Gamma=4/3$ polytrope, because they only consider low-degree modes.} Then, the corrected frequencies were compared to the ``exact'' ones, obtained from complete simulations. What they found is that second-order corrections are satisfactory for the high-frequency $g$-modes (low overtones), but even third-order corrections become insufficient very early for the low-frequency ones. They attribute this result to the fact that, in the subinertial regime $(\omega<2\Omega)$, the modes acquire a mixed gravito-inertial character \cite{UnnoEtAl1989}, which significantly changes their propagation zone \cite{DintransRieutord2000}, a property which 
is not considered by perturbative methods.

Similar calculations for $p$-modes (acoustic modes) have also been performed \cite{LignieresEtAl2006,ReeseEtAl2006}, where it was shown that the slow-rotation approximation, even at third-order, fails  at relatively low rotation rates. However, $p$-modes reside at a frequency range which is too high for our resonance condition to be satisfied. As mentioned in \cref{sec:Introduction} and thoroughly explained in Paper I, the coupling occurs only if the parent mode frequency $\omega_\alpha$ nearly equals the daughter mode frequencies $\omega_\beta+\omega_\gamma$. As a result, only modes with frequencies lower than the $f$-mode frequency can become suitable daughters, hence $g$-modes and CFS-stable $f$-modes ($>1500$ modes).

Given the approximations we have applied, we immediately see that the couplings among the modes can significantly change if we consider their ``correct'' eigenfrequencies and eigenfunctions. A daughter pair that resonates with the parent in the slow-rotation approximation might not do so in the complete solution. Furthermore, the strength of the various couplings should be affected by the form of the eigenfunctions, which, at the large rotation rates considered here, are expected to differ from their nonrotating counterparts (e.g., see \refcite{KastaunEtAl2010}). However, the nature of the problem is such that a precise evaluation of the coupled triplet network is not the important point. What we are looking for is a low-order estimation of the value of the lowest PIT, around which the parent saturates. Since the daughter pair responsible for the saturation of the parent is chosen from a ``sea'' of available modes, this a highly statistical process, from which the triplet that minimizes \cref{PIT} is always picked. Besides, even if we could have used the exact eigenfrequency and eigenfunction solutions, we would, at best, have calculated the correct couplings of a very simple neutron star model, rife with other simplifications and approximations.

While the slow-rotation approximation is used for the daughter modes, the same cannot be done for the parent modes. The $f$-mode instability becomes active at large angular velocities, close to the Kepler (mass-shedding) limit $\Omega_\mathrm{K}$ and, as a result, even second-order rotational corrections do not suffice for the $f$-modes to become unstable. To fix this, we manually introduced ``higher-order'' corrections to their frequencies, based on the exact solutions provided in \refcite{IpserLindblom1990}\footnote{In particular, we introduced third- and fourth-order terms in \cref{rotationally corrected eigenfrequency}, so that parent mode eigenfrequencies fit the curves and values given in Figs.~2--4 and Table 1 of \refcite{IpserLindblom1990}.}---alternatively, one could use the empirical frequency relations in \refcite{DonevaEtAl2013,DonevaKokkotas2015}. %(even though they were extracted for relativistic stars with realistic equations of state)

We should note that, in principle, coupling of the $f$-mode to inertial modes (like the $r$-mode) can be possible, as it is not forbidden by any coupling selection rule \cite{SchenkEtAl2001}. The main reason we only considered polar modes is that the coupling coefficient for polar mode coupling has a relatively simple, known form \cite{Dziembowski1982}, which was presented in Paper I. By considering a stratified star ($\Gamma\ne\Gamma_1$, with $\Gamma_1\approx\mathrm{\textit{const.}}$, which is the simplest case), we get the low-frequency modes that will play the role of daughter modes, i.e., $g$-modes. We could have cases where an $r$-mode is one of the daughter modes but, since $r$-modes are purely axial (to zeroth order in $\Omega$) in stratified stars \cite{PapaloizouPringle1978}, this would make the coupling less efficient. If the coupling coefficient were evaluated to higher orders in the rotation, coupling to $r$-modes could become significant. Had we considered an isentropic star $(\Gamma=\Gamma_1)$, then no $g$-modes would be present (more precisely, they would become trivial). In this case, the daughters would have to be CFS-stable $f$-modes and generalized $r$-modes. The latter have also 
been called hybrid rotational modes, because, in the nonrotating limit, they have both polar and axial components \cite{LindblomIpser1999,LockitchFriedman1999}. This could make them more suitable daughters than the ``classical'' $r$-modes, but would also require modifications in the form of the coupling coefficient, accounting for the additional axial components of the daughters.

As briefly mentioned before though, $g$-modes are also driven by rotation, together with buoyancy, in rotating configurations \cite{DintransRieutord2000,YoshidaLee2000} and have been shown to approach the rotational modes of isentropic stars for large angular velocities \cite{Passamonti2009,PassamontiEtAl2009,GaertigKokkotas2009}. Given that the $f$-mode instability operates at high rotation rates, this suggests that either studying stratified or nonstratified stars would not affect the coupling, since, for both cases, all the daughter modes (except for the CFS-stable $f$-modes) would be of the inertial type. In practice, however, the slow-rotation approximation that we use does not take into account this inertial-led behavior of $g$-modes, even though their frequencies are dominated by the correction terms, for large $\Omega$.

%-----------------------------------------------------------------------------------------------------------------------------%
%-----------------------------------------------------------------------------------------------------------------------------%

\subsection{Growth/damping rates} \label{subsec:Growth/damping rates}

For the growth $(\gamma>0)$ or damping $(\gamma<0)$ rates of the modes we consider the basic mechanisms for the dissipation of fluid oscillations, of a neutron star consisting of normal nuclear matter, namely (nonsuperfluid) neutrons, (nonsuperconducting) protons, and electrons. These are gravitational waves (GW), bulk viscosity (BV), and shear viscosity (SV). All the relevant formulas for this section can be found in Paper I.

Gravitational radiation damps the mode, unless $\omega(\omega-m\Omega)<0$, where $\omega$ denotes the corotating frame frequency and $\omega-m\Omega$ the inertial frame frequency of the mode. In other words, a sign change of the inertial frame frequency signifies the onset of the CFS instability for the mode. Shear and bulk viscosity act against it, damping the mode at low $(\lesssim 10^9\,\mathrm{K})$ and high $(\gtrsim 10^9\,\mathrm{K})$ temperatures respectively, leaving a small region in the $(T,\Omega)$ plane where the instability occurs (instability window). This region is defined by
\begin{equation}
	\gamma=\gamma_\mathrm{GW}+\gamma_\mathrm{BV}+\gamma_\mathrm{SV}>0. \label{instability condition}
\end{equation}
The time scale over which the instability grows is, by definition, $\tau=1/\gamma$.

The low $f$-mode $l=m$ multipoles are the most unstable due to the CFS mechanism \cite{IpserLindblom1990,*IpserLindblom1991}. Low $g$-mode overtones might also be susceptible to the instability \cite{Lai1999,PassamontiEtAl2009}, albeit with much longer time scales, which probably makes them unimportant to the evolution of the star.

All the growth/damping rates are evaluated in the nonrotating limit, with the exception of the factor $\omega(\omega-m\Omega)$, which is calculated using the corrected eigenfrequencies. Use of the rotationally corrected eigenfunctions spoils the direct spherical harmonic dependence of the mode, making the evaluation of the various quantities harder to follow. Therefore, only $\gamma_\mathrm{GW}$ changes with $\Omega$, whereas $\gamma_\mathrm{BV}$ and $\gamma_\mathrm{SV}$ depend solely on the temperature, scaling as $T^6$ and $T^{-2}$ respectively.

%-----------------------------------------------------------------------------------------------------------------------------%
%-----------------------------------------------------------------------------------------------------------------------------%

\subsection{Couplings} \label{subsec:Couplings}

Having calculated all the quantities associated with every mode, namely its eigenfrequency, its eigenfunction, and its growth/damping rate, we proceed with finding all the possible couplings between unstable $f$-modes and the rest of the polar modes considered. We choose the $l=m=2,\,3,\,4$ $f$-modes to be the parent modes, because they have the best ``instability window size/growth time scale'' ratio among all the unstable polar modes.

We subject all the possible parent-daughter-daughter triplets to a screening process, using the coupling selection rules as criteria. Although the selection rules for the orders $m_i$ and the degrees $l_i$ are either satisfied or not (see Paper I), there is an inherent freedom in the resonance condition, stemming from the detuning parameter $\Delta\omega$. Thus, we define a cutoff parameter $\Delta\omega_\mathrm{\,max}$ such that, if $|\Delta\omega|\le\Delta\omega_\mathrm{\,max}$, then the modes are considered resonant. The actual value of this parameter is chosen after a few trial runs, so that the triplets with the lowest PITs do not change by further increasing it.

Then, we proceed with the calculation of the coupling coefficient $\mathcal{H}$ for every coupled triplet. As mentioned before, we are evaluating the coupling coefficient in the nonrotating limit, using Eq.~(B3) from Paper I. The angular dependence of the coupling coefficient is thus reduced to a simple spherical harmonic integral \cite{Dziembowski1982}, which does not happen if one considers the rotationally corrected eigenfunctions instead.

The value of the coupling coefficient is normalization-dependent, together with the values of the amplitudes $Q_i$ in \cref{equations of motion with normalisation choice}. Since the energy of a mode in the corotating frame, given by $E_i=|Q_i|^2 E_\mathrm{unit}$ \cite{SchenkEtAl2001}, should be normalization-independent, we can rewrite \cref{equations of motion with normalisation choice} as
\begin{subequations}
	\label[subequations]{equations of motion normalisation-independent}
	\begin{eqnarray}
		\dot{\mathscr{Q}}_\alpha &=& \gamma_\alpha \mathscr{Q}_\alpha+i\omega_\alpha\mathscr{H} \, \mathscr{Q}_\beta \mathscr{Q}_\gamma e^{-i\Delta\omega t}, \label{mode alpha equation of motion normalisation-independent} \\
		\dot{\mathscr{Q}}_\beta &=& \gamma_\beta \mathscr{Q}_\beta+i\omega_\beta\mathscr{H} \, \mathscr{Q}_\gamma^* \mathscr{Q}_\alpha e^{i\Delta\omega t}, \label{mode beta equation of motion normalisation-independent} \\
		\dot{\mathscr{Q}}_\gamma &=& \gamma_\gamma \mathscr{Q}_\gamma+i\omega_\gamma\mathscr{H} \, \mathscr{Q}_\alpha \mathscr{Q}_\beta^* e^{i\Delta\omega t}, \label{mode gamma equation of motion normalisation-independent}
	\end{eqnarray}
\end{subequations}
where $\mathscr{Q}_i=Q_i E_\mathrm{unit}^{1/2}$ and $\mathscr{H}=\mathcal{H}/E_\mathrm{unit}^{3/2}$ are normalization-independent quantities. So, for another normalization choice $E'_\mathrm{unit}$, the coupling coefficient transforms as $\mathcal{H'}/\mathcal{H}=(E'_\mathrm{unit}/E_\mathrm{unit})^{3/2}$.

Now, we are finally able to calculate every coupled triplet's PIT, using \cref{PIT}.

%-----------------------------------------------------------------------------------------------------------------------------%
%-----------------------------------------------------------------------------------------------------------------------------%

\subsection{Saturation} \label{subsec:Saturation}

For the last step, all the coupled triplets are sorted in ascending order, according to their PITs. Starting with the triplet that has the lowest PIT, we can examine whether it leads to saturation or not.

For a daughter pair to successfully stop the parent's growth and make it saturate two conditions have to be fulfilled, given by
\begin{equation}
	|\gamma_\beta+\gamma_\gamma| \gtrsim \gamma_\alpha \label{saturation condition 1}
\end{equation}
and
\begin{equation}
	|\Delta\omega| \gtrsim |\gamma_\alpha+\gamma_\beta+\gamma_\gamma|. \label{saturation condition 2}
\end{equation}
These relations are approximate representations of Eq.~(4.24) in Paper I, whose derivation is based on a linear stability analysis of the equilibrium solution of \cref{equations of motion with normalisation choice}.

If these conditions are met for the daughter pair with the lowest PIT, the triplet's amplitudes successfully converge towards their equilibrium solution. If not, the parent will keep growing, albeit more slowly, until the next PIT is crossed and a different daughter pair is excited. The equilibrium solution of the \emph{new} triplet is now examined for stability [through \cref{saturation condition 1,saturation condition 2}]. This process continues, until the first stable equilibrium is found. The lowest PIT that leads to successful saturation will be called \emph{stable} and approximately equals the parent's saturation amplitude (see Paper I).

Usually, the triplet with the lowest PIT does satisfy the saturation conditions. In the few cases where it does not, the picture described above can actually get more convoluted. An overview of the influence of the two conditions on the coupled-mode system, together with a more detailed description of the saturation mechanism, is presented in \cref{sec:The saturation mechanism}.

%-----------------------------------------------------------------------------------------------------------------------------%
%%%%%%%%%%%%%%%%%%%%%%%%%%%%%%%%%%%%%%%%%%%%%%%%%%%%%%%%%%%%%%%%%%%%%%%%%%%%%%%%%%%%%%%%%%%%%%%%%%%%%%%%%%%%%%%%%%%%%%%%%%%%%%%
%-----------------------------------------------------------------------------------------------------------------------------%

\section{Results} \label{sec:Results}

%-----------------------------------------------------------------------------------------------------------------------------%
%-----------------------------------------------------------------------------------------------------------------------------%

\subsection{Supernova-derived neutron stars} \label{subsec:Supernova-derived neutron stars}

%-----------------------------------------------------------------------------------------------------------------------------%

\subsubsection{Instability evolution}

The $f$-mode instability is expected to be important in newborn neutron stars \cite{PassamontiEtAl2013}. After a core-collapse supernova explosion, a very hot $(T\sim 10^{11}\,\mathrm{K})$ proto-neutron star is formed \cite{BurrowsLattimer1986,Burrows1990}, which subsequently cools down due to neutrino emission, powered by the so-called Urca processes (e.g., \refcite{PrakashEtAl2001}). Depending on its angular velocity, the star might enter the $f$-mode instability window. For the Newtonian polytropes that we use, this means that we require initial angular velocities $\Omega>0.9\,\Omega_\mathrm{K}$ (relativity and realistic equations of state can drag the window down to $0.8\,\Omega_\mathrm{K}$ \cite{DonevaEtAl2013,GaertigEtAl2011}).

Conservation of angular momentum during the core-collapse phase makes spin periods close to the Kepler limit theoretically feasible ($\sim 1\,\mathrm{ms}$; e.g., \refcite{HegerEtAl2000}). Observations of young pulsars, however, imply initial rotation periods $\sim 10-100\,\mathrm{ms}$, suggesting the involvement of some mechanism which either spins down the star at an early stage in its life (see \refcite{OttEtAl2006} for such possible mechanisms) or makes it spin slowly from the outset. Hence, the $f$-mode instability scenario is relevant for neutron stars which rotate fast after their birth.

As soon as the star enters the instability window, the unstable mode grows exponentially until it saturates. Shear viscosity, triggered by the oscillation, heats up the star and balances neutrino cooling, establishing thermal equilibrium.\footnote{As opposed to shear viscosity, bulk viscosity cools down the star by neutrino emission. However, its contribution to the star's cooling is negligible \cite{PassamontiEtAl2013}.} Gravitational waves emitted from the perturbed star carry off angular momentum and the star descends the instability window along a thermal equilibrium curve $(T\approx\mathrm{\textit{const.}})$, until it finally exits the window (see \refcite{PassamontiEtAl2013} for the $f$-mode evolution in nascent neutron stars, and \refcite{OwenEtAl1998,BondarescuEtAl2007,*BondarescuEtAl2009} and \cite{AnderssonEtAl2002} for the $r$-mode evolution in nascent neutron and strange stars respectively).

The saturation amplitude of the unstable $f$-mode determines the gravitational wave strain associated with the perturbation. The detectability of the signal also depends on the competition of the $f$-mode instability with other spin-down mechanisms, such as the $r$-mode instability and magnetic braking. Should the $r$-mode saturation amplitude be larger than (or even comparable to) the $f$-mode one, or the magnetic field be greater than some critical value, then one or both of these mechanisms will dominate the spin evolution of the neutron star \cite{PassamontiEtAl2013}.

%-----------------------------------------------------------------------------------------------------------------------------%

\subsubsection{Models}

To study the $f$-mode saturation, we applied the (Newtonian) formalism presented in Paper I in polytropic stars. We used two polytropic configurations, with $\Gamma=2$ and $3$, and varied the adiabatic exponent $\Gamma_1$, leading to strongly or weakly stratified stars (the smaller the difference between $\Gamma_1$ and $\Gamma$, the closer to zero the $g$-mode frequencies are pushed in the nonrotating limit; see Paper I). Because of the complications described in \cref{subsec:Eigenfrequencies and eigenfunctions} regarding $g$-mode frequencies, models in which $\Gamma_1-\Gamma$ was very small exhibited divergent behavior and thus were ignored.

\begin{figure*}
	\includegraphics[width=.55\textwidth,keepaspectratio]{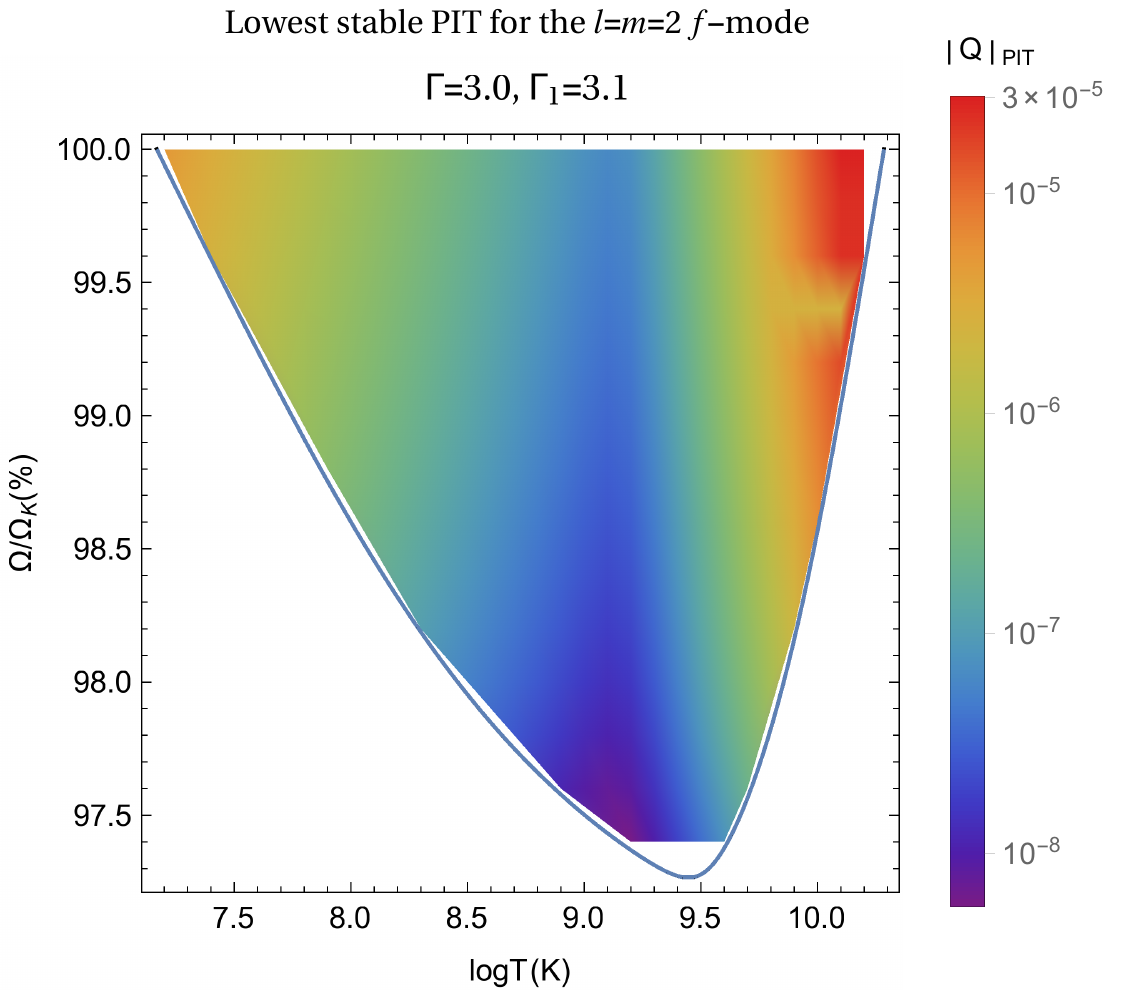}
	\caption{Contour plot of the lowest stable PIT ($\approx$ saturation amplitude) inside the instability window of the $l=m=2$ $f$-mode, for a supernova-derived neutron star with $M=1.4\,M_\odot$ and $R=10\,\mathrm{km}$. The star is described by a polytrope with $\Gamma=3$ and an adiabatic exponent $\Gamma_1=3.1$. The mode amplitude is given by the relation $|Q|=\sqrt{E_\mathrm{mode}/E_\mathrm{unit}}$, with $E_\mathrm{unit}=Mc^2$.}
	\label{fig:2f_PIT_SDNS}
\end{figure*}

\begin{sidewaysfigure*}
	\centering
	\includegraphics[width=\textwidth,keepaspectratio]{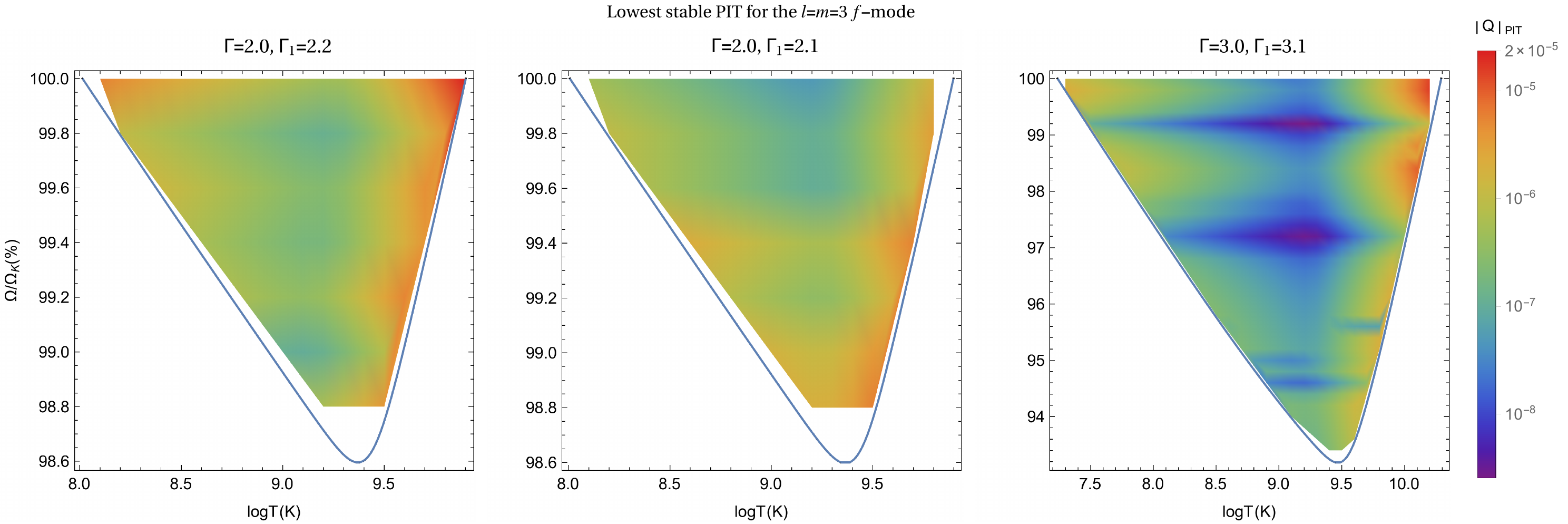}
	\caption{Contour plots of the lowest stable PIT ($\approx$ saturation amplitude) inside the instability window of the $l=m=3$ $f$-mode, for a supernova-derived neutron star with $M=1.4\,M_\odot$ and $R=10\,\mathrm{km}$. The star is described by a polytrope with $\Gamma=2$, and an adiabatic exponent $\Gamma_1=2.2$ and $2.1$, as well as a polytrope with $\Gamma=3$ and $\Gamma_1=3.1$. The mode amplitude is given by the relation $|Q|=\sqrt{E_\mathrm{mode}/E_\mathrm{unit}}$, with $E_\mathrm{unit}=Mc^2$.}
	\label{fig:3f_PIT_SDNS}
\end{sidewaysfigure*}

\begin{sidewaysfigure*}
	\centering
	\includegraphics[width=\textwidth,keepaspectratio]{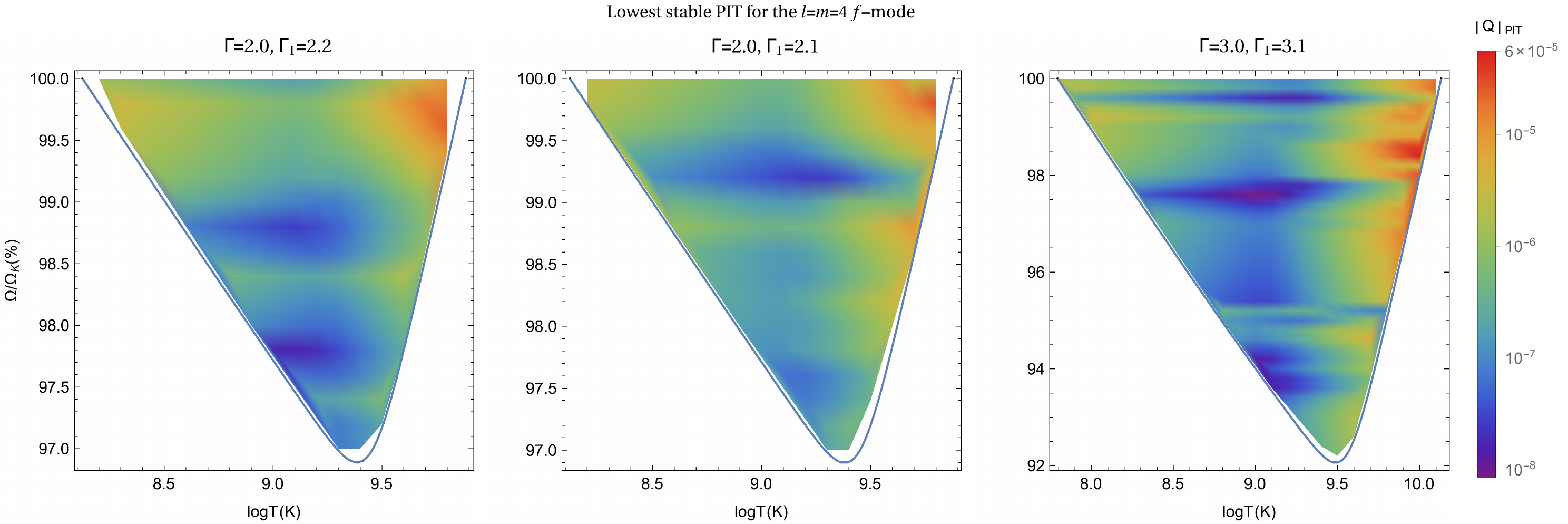}
	\caption{Contour plots of the lowest stable PIT ($\approx$ saturation amplitude) inside the instability window of the $l=m=4$ $f$-mode, for a supernova-derived neutron star with $M=1.4\,M_\odot$ and $R=10\,\mathrm{km}$. The star is described by a polytrope with $\Gamma=2$, and an adiabatic exponent $\Gamma_1=2.2$ and $2.1$, as well as a polytrope with $\Gamma=3$ and $\Gamma_1=3.1$. The mode amplitude is given by the relation $|Q|=\sqrt{E_\mathrm{mode}/E_\mathrm{unit}}$, with $E_\mathrm{unit}=Mc^2$.}
	\label{fig:4f_PIT_SDNS}
\end{sidewaysfigure*}

The results for three models are presented in \cref{fig:2f_PIT_SDNS,fig:3f_PIT_SDNS,fig:4f_PIT_SDNS}, where we plot the lowest stable PIT ($\approx$ saturation amplitude, see \cref{subsec:Saturation}) throughout the instability window. In the first two models, $\Gamma=2$, and $\Gamma_1=2.2$ and $2.1$, whereas in the third one $\Gamma=3$ and $\Gamma_1=3.1$. The unstable $f$-modes we consider are the quadrupole ($l=m=2$), the octupole ($l=m=3$), and the hexadecapole ($l=m=4$; see \cref{subsec:Couplings}). All three of these modes become unstable in the $\Gamma=3$ polytrope, but only the last two in the $\Gamma=2$ polytrope.

In all three models $M=1.4\,M_\odot$ and $R=10\,\mathrm{km}$, where $M_\odot$ is the solar mass. The normalization used is $E_\mathrm{unit}=Mc^2$, which is the mode energy at unit amplitude $(E_\mathrm{mode}=|Q|^2 E_\mathrm{unit})$.

\begin{figure*}
	\includegraphics[width=\textwidth,keepaspectratio]{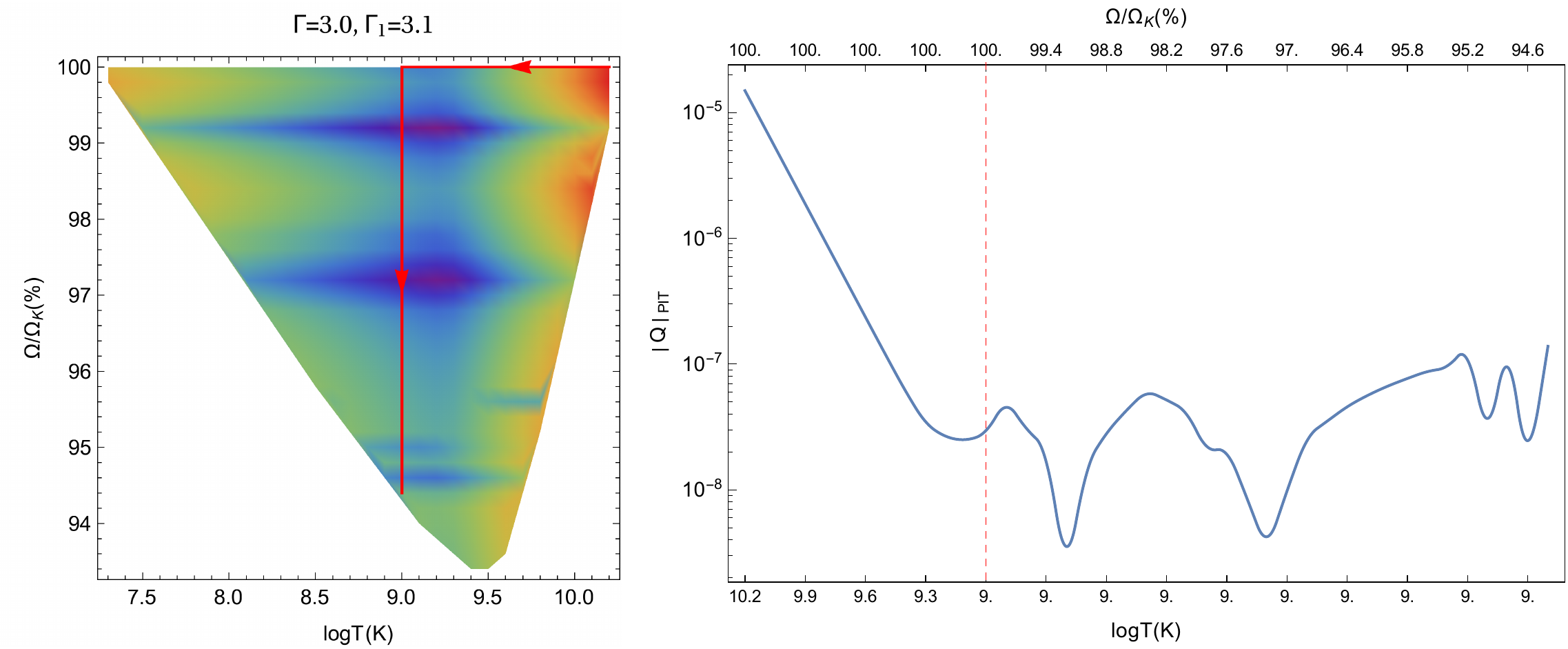}
	\caption{Hypothetical evolution of a supernova-derived neutron star with $M=1.4\,M_\odot$ and $R=10\,\mathrm{km}$, through the instability window of the $l=m=3$ $f$-mode (\textit{left}), and the corresponding evolution of the lowest stable PIT (\textit{right}). The star obeys a polytropic equation of state with $\Gamma=3$ and an adiabatic exponent $\Gamma_1=3.1$. In this example, the star enters the window during its cooling phase, rotating at its maximum angular velocity, until thermal equilibrium is established (indicated by the vertical dashed line), at which point it descends the window at $T=10^9\,\mathrm{K}$ \cite{PassamontiEtAl2013}.}
	\label{fig:3f_PIT_SDNS_hypothetical_evolution}
\end{figure*}

In addition, a hypothetical (based on the results of \refcite{PassamontiEtAl2013}) evolution of a star inside the instability window of the octupole $f$-mode, together with the variation of the mode's saturation amplitude, can be seen in \cref{fig:3f_PIT_SDNS_hypothetical_evolution}.

%-----------------------------------------------------------------------------------------------------------------------------%

\subsubsection{Discussion}

Two main features can be observed in \cref{fig:2f_PIT_SDNS,fig:3f_PIT_SDNS,fig:4f_PIT_SDNS}: a) the decrease of the saturation amplitude from the edge to the interior of the instability window, and b) horizontal bands where the amplitude behaves differently from the ``background''---although the bands themselves also follow the modulation of the first feature.

\hypertarget{First feature}{\paragraph{First feature.}} The first feature can be clearly seen in \cref{fig:2f_PIT_SDNS}, where the second feature is absent. This decline of the saturation amplitude can easily be explained if one looks at the coupling spectrum, i.e., the daughter pairs responsible for the saturation of the parent throughout the instability window.

As a rule, we have two types of daughter pairs: either a CFS-stable $f$-mode and a $g$-mode ($f$-$g$ coupling) or two $g$-modes ($g$-$g$ coupling). Depending on the coupling type and the daughters' parameters, we can simplify the formula for the PIT \eqref{PIT}, as shown in \cref{sec:Approximate relations for the parametric instability threshold}.

In the case of $f$-$g$ couplings, which is the most common, the $f$-mode damping rate $\gamma_\beta$ is much larger (in absolute value) than the $g$-mode damping rate $\gamma_\gamma$. Then, \cref{PIT} is approximated by \cref{PIT gamma_beta>>gamma_gamma Deltaomega=gamma_beta} or \eqref{PIT gamma_beta>>gamma_gamma Deltaomega>>gamma_beta}. Since the $f$-mode damping is mainly due to gravitational waves, it does not change much with temperature, which makes it roughly constant for some angular velocity $\Omega$. On the other hand, for this type of coupling, the $g$-mode daughter is predominantly damped by viscosity. From \cref{PIT gamma_beta>>gamma_gamma Deltaomega=gamma_beta,PIT gamma_beta>>gamma_gamma Deltaomega>>gamma_beta}, this means that
\begin{equation}
	|Q_\mathrm{PIT}|\propto\sqrt{|\gamma_\gamma|}\propto
	\left\{
	\begin{array}{ll}
		T^{-1}, & T\lesssim 10^9\,\mathrm{K} \\
		T^3, & T\gtrsim 10^9\,\mathrm{K}
	\end{array}
	\right.
	\;\mathrm{for}\;\;\Omega=\mathrm{\textit{const.}}
	\label{f-g coupling saturation amplitude scaling}
\end{equation} 
In other words, along an $\Omega=\mathrm{\textit{const.}}$ line, the saturation amplitude follows the behavior of the $g$-mode daughter damping rate.\footnote{The daughters' damping rates $\gamma_\beta$ and $\gamma_\gamma$ are the only quantities in \cref{PIT} that can change along a constant angular velocity line.} The temperature dependence is a result of shear and bulk viscosity, dominating the damping at low and high temperatures, and scaling as $T^{-2}$ and $T^6$, respectively (see \cref{subsec:Growth/damping rates}).

When a $g$-$g$ coupling prevails, the relation between the daughters' damping rates can vary. If gravitational waves dominate the damping for one of them, everything is reduced to the $f$-$g$ coupling case. This happens when one of them is a low $g$-mode multipole. If viscosity dominates for both, it is better to start with the observation that the detuning is usually much larger (in absolute value) than the damping rates. Then, the relevant approximate formulas for the PIT are \cref{PIT gamma_beta>>gamma_gamma Deltaomega>>gamma_beta,PIT gamma_beta=gamma_gamma Deltaomega>>gamma_beta}. Both equations show that the saturation amplitude should not change along constant angular velocity lines. This is obvious in \cref{PIT gamma_beta=gamma_gamma Deltaomega>>gamma_beta}, but can also be seen in \cref{PIT gamma_beta>>gamma_gamma Deltaomega>>gamma_beta}, because $|Q_\mathrm{PIT}|\propto\sqrt{\gamma_\gamma/\gamma_\beta}$ and the damping rates follow the same temperature scaling. Nevertheless, this situation is not observed much (see 
paragraph \hyperlink{Is the number of modes enough?}{\textit{d}} below).

% Original: The reason that the latter situation is not observed much is because such daughter pairs, mainly damped by viscosity, only marginally satisfy the saturation condition \eqref{saturation condition 1}. As a result, they might sometimes give the lowest stable PIT near the edge of the window, where viscosity damping is large enough, but fail to saturate the unstable mode at intermediate temperatures, in which case other triplets, with larger damping rates, will take their place. Update: it is basically not observed much, because, when $\Delta\omega\gg\gamma_\beta$, $|Q|^2\propto\Delta\Omega^2$. This does not give as low amplitudes as the case when $\Delta\omega\approx\gamma_\beta$, where $|Q|^2\propto\gamma_\beta\gamma_\gamma$, because typically $\Delta\omega^2\gg\gamma_\beta\gamma_\gamma$.

\paragraph{Second feature.} For the second feature to be understood, we need to look at constant temperature lines instead. The difficulty here is that all quantities that appear in \cref{PIT} change as $\Omega$ is varied. This makes the modulation of the saturation amplitude along a $T=\mathrm{\textit{const.}}$ line harder to follow.

Looking at the coupling spectrum, we see that the same daughter pair is usually responsible for the saturation of the parent along an $\Omega=\mathrm{\textit{const.}}$ line. After all, this is the basis of the reasoning that led to \cref{f-g coupling saturation amplitude scaling}. This is no longer true along constant temperature lines: the daughter pair which gives the lowest stable PIT may change many times. Occasionally, this change might be abrupt, making the saturation amplitude higher or lower, compared to neighboring angular velocities. As a result, these characteristic horizontal bands appear, which, however, individually still follow the behavior of the first feature.

Although the effect is highly statistical, given the number of variables and available modes (see \cref{subsec:Eigenfrequencies and eigenfunctions}), we can single out two main reasons for it: The first is the occurrence of a very fine resonance between the parent and some daughter pair, which only appears for a specific angular velocity. Such a resonance has a very low detuning $|\Delta\omega|$, which can lead to the drop of the saturation amplitude. The second, less frequent, reason is related to the validity of the saturation conditions \eqref{saturation condition 1} and \eqref{saturation condition 2}. If the saturating triplet satisfies one of these conditions marginally for some value of the angular velocity, it will not be long before it cannot saturate the parent any more, and some other daughter pair will take its place.

\paragraph{Varying the adiabatic exponent.} As mentioned before, using different values for the difference between the adiabatic and polytropic exponents, $\Gamma_1-\Gamma$, shifts the nonrotating-limit $g$-mode frequencies closer to or further away from the $f$-mode frequency (the latter depends mainly on $\Gamma$ and is highly unaffected by any change in $\Gamma_1$). However, since we are interested in fast-rotating stars, rotational corrections to $g$-mode frequencies will prevail, causing $g$-modes to become rotationally-driven, rather than buoyancy-driven (see \cref{subsec:Eigenfrequencies and eigenfunctions}).

Consequently, fast-rotating models with different adiabatic exponents (but the same polytropic exponent) should be nearly indistinguishable---at least as far as $f$- and $g$-modes are concerned. This means that the couplings and the saturation amplitudes should not change much if a different value of $\Gamma_1$ is chosen for some polytrope. This can indeed be seen, to some extent, in \cref{fig:3f_PIT_SDNS,fig:4f_PIT_SDNS}. In practice, however, as discussed in \cref{subsec:Eigenfrequencies and eigenfunctions}, $g$-modes do not exhibit inertial behavior in the slow-rotation approximation. Hence, in principle, there should be differences in the results if one considers coupling to inertial modes; for instance, the inertial mode damping rates reported by \refcite{LockitchFriedman1999} are larger (in absolute value) than our $g$-mode damping rates, which, according to \cref{f-g coupling saturation amplitude scaling}, should systematically \emph{increase} the saturation amplitude.

\hypertarget{Is the number of modes enough?}{\paragraph{Is the number of modes enough?}} In our models, we searched for couplings of unstable $f$-modes to more than 1500 polar modes and obtained many triplets with fine resonances, meaning that our frequency spectrum was dense enough for the parent to always resonate with daughter pairs. These fine resonances could probably become even finer and/or more frequent, had we included more modes in the calculation. However, a small detuning alone does not necessarily lead to smaller amplitudes. This can be seen in many $g$-$g$ couplings, where even though better resonances were achieved compared to $f$-$g$ couplings, the latter were much more abundant in the coupling spectrum (see paragraph \hyperlink{First feature}{\textit{a}} above). This shows that triplets with larger detunings might give the lowest PITs instead, depending on how their parameters are tuned and which of the cases shown in \cref{sec:Approximate relations for the parametric instability threshold} 
they fall under. 

%-----------------------------------------------------------------------------------------------------------------------------%
%-----------------------------------------------------------------------------------------------------------------------------%

\subsection{Merger-derived neutron stars} \label{subsec:Merger-derived neutron stars}

%-----------------------------------------------------------------------------------------------------------------------------%

\subsubsection{Instability evolution}

As briefly mentioned in \cref{sec:Introduction}, the $f$-mode instability could also be important in post-merger neutron star remnants \cite{DonevaEtAl2015}. After the binary coalescence, a supramassive neutron star may be formed, supported by rotation against collapse \cite{CookEtAl1992,*CookEtAl1994} (see also simulations in \refcite{HotokezakaEtAl2013,KastaunGaleazzi2015}). Such stars have been associated with $\gamma$-ray bursts in an attempt to explain their persistent afterglow activity \cite{DaiLu1998,ZhangMeszaros2001,RowlinsonEtAl2013} (see also \refcite{CiolfiSiegel2015,RezzollaKumar2015}). According to Fig.~1 in \refcite{DonevaEtAl2015}, the instability growth time scale due to gravitational waves, $\tau_\mathrm{GW}$, can be as short as $10-100\,\mathrm{s}$ for $M>2.4\,M_\odot$ (for typical neutron stars, this time scale increases by orders of magnitude). However, the lifetimes of supramassive stars are very limited, because, as they spin down, a point is reached when centrifugal support can no longer prevent gravitational collapse.

Thus, apart from rotating fast enough, supramassive stars should also survive for enough time, for the $f$-mode instability to develop. This is fairly supported by recent calculations \cite{RaviLasky2014}, which suggest that these objects may remain stable for up to $\approx 4\times 10^4\,\mathrm{s}$. After the initial differentially rotating and cooling phase, the star might enter the instability window and follow a path similar to the one described in \cref{subsec:Supernova-derived neutron stars} for supernova-derived neutron stars, but, since the window for a supramassive star is quite larger, the star may collapse to a black hole before it exits the window \cite{DonevaEtAl2015}.

As in the case of supernova-derived neutron stars, the saturation amplitude of the unstable $f$-mode determines whether the associated gravitational wave signal can be detected, based on the competition among the $f$-mode instability, the $r$-mode instability, and magnetic braking (see Figs.~2 and 3 in \refcite{DonevaEtAl2015}).

%-----------------------------------------------------------------------------------------------------------------------------%

\subsubsection{Models}

We applied the (Newtonian) mode coupling formalism in configurations which \emph{emulate} merger-derived, supramassive stars.\footnote{Such stars do not actually admit a Newtonian limit \cite{CookEtAl1992}.} We considered a star with $M=2.5\,M_\odot$ and $R=12\,\mathrm{km}$, obeying a polytropic equation of state with $\Gamma=3$, and an adiabatic exponent $\Gamma_1=3.2$ and $3.1$.

In order to achieve the instability growth time scales of \refcite{DonevaEtAl2015}, we manually enhanced the Kepler limit of our models.\footnote{This way, the factor $\omega(\omega-m\Omega)$, appearing in the gravitational-wave growth rate formula (see \cref{subsec:Growth/damping rates}), can obtain larger (absolute) values.} These time scales cannot be obtained legitimately by our Newtonian polytropes, because the models used in \refcite{DonevaEtAl2015} are relativistic (employing the Cowling approximation) and governed by realistic equations of state.

Given the assumptions above and the simplicity of our approach, our models should be merely considered as toy models, used to demonstrate the impact of larger masses and shorter instability growth times on the saturation amplitude of the unstable modes.

\begin{sidewaysfigure*}
	\centering
	\includegraphics[width=\textwidth,keepaspectratio]{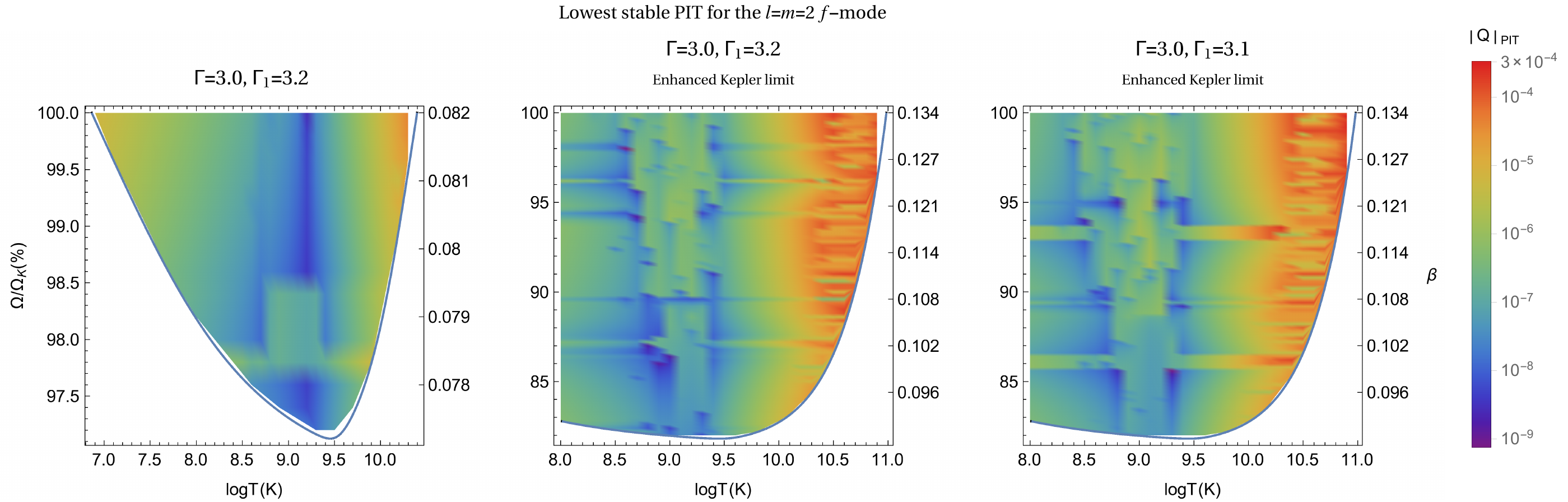}
	\caption{Contour plots of the lowest stable PIT ($\approx$ saturation amplitude) inside (part of) the instability window of the $l=m=2$ $f$-mode, for our toy model of a merger-derived, supramassive neutron star with $M=2.5\,M_\odot$ and $R=12\,\mathrm{km}$. The star is described by a polytrope with $\Gamma=3$, and an adiabatic exponent $\Gamma_1=3.2$ and $3.1$. The Kepler limit has been enhanced, to imitate the behavior of the models used in Ref.~\cite{DonevaEtAl2015}. A model with its actual Kepler limit is also shown. The mode amplitude is given by the relation $|Q|=\sqrt{E_\mathrm{mode}/E_\mathrm{unit}}$, with $E_\mathrm{unit}=Mc^2$.}
	\label{fig:2f_PIT_MDNS}
\end{sidewaysfigure*}

\begin{sidewaysfigure*}
	\centering
	\includegraphics[width=\textwidth,keepaspectratio]{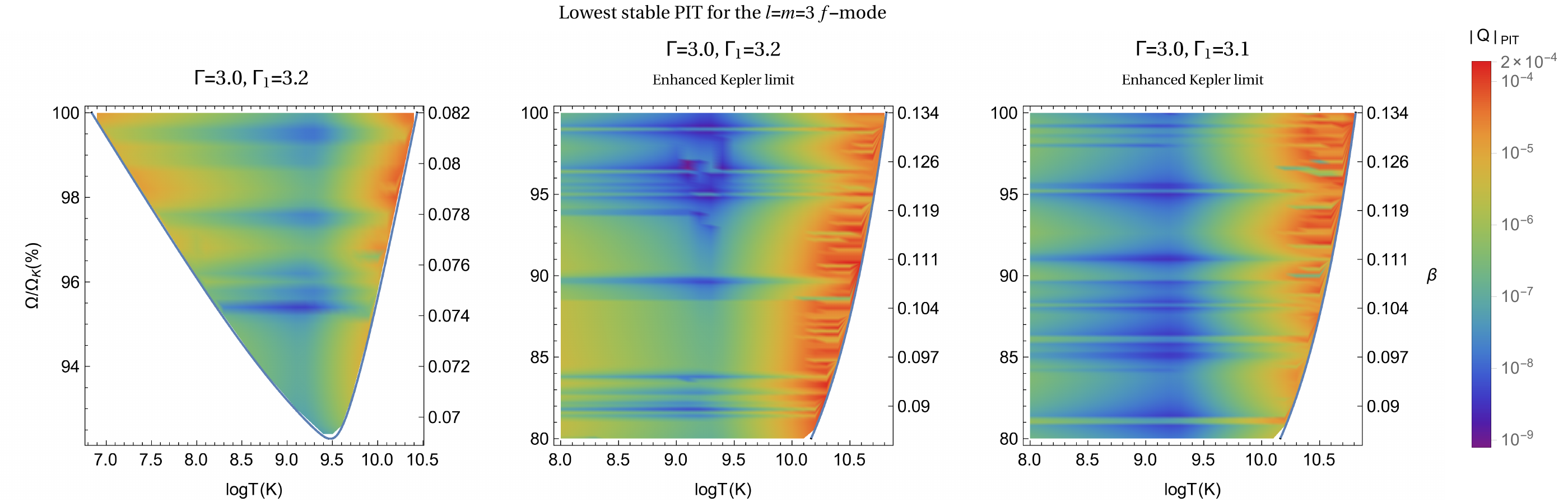}
	\caption{Contour plots of the lowest stable PIT ($\approx$ saturation amplitude) inside (part of) the instability window of the $l=m=3$ $f$-mode, for our toy model of a merger-derived, supramassive neutron star with $M=2.5\,M_\odot$ and $R=12\,\mathrm{km}$. The star is described by a polytrope with $\Gamma=3$, and an adiabatic exponent $\Gamma_1=3.2$ and $3.1$. The Kepler limit has been enhanced, to imitate the behavior of the models used in Ref.~\cite{DonevaEtAl2015}. A model with its actual Kepler limit is also shown. The mode amplitude is given by the relation $|Q|=\sqrt{E_\mathrm{mode}/E_\mathrm{unit}}$, with $E_\mathrm{unit}=Mc^2$.}
	\label{fig:3f_PIT_MDNS}
\end{sidewaysfigure*}

The results for the models described above are presented in \cref{fig:2f_PIT_MDNS,fig:3f_PIT_MDNS}, where the lowest stable PIT ($\approx$ saturation amplitude, see \cref{subsec:Saturation}) is plotted inside the instability window of the quadrupole $(l=m=2)$ and the octupole $(l=m=3)$ $f$-modes. The normalization used for the mode energy is $E_\mathrm{unit}=Mc^2$. Since the left part of the windows is not expected to be significant for the evolution of a newborn neutron star, and given their considerably larger size, compared to the corresponding windows from supernova-derived stars, we restricted our calculations to $T\ge 10^8\,\mathrm{K}$. Furthermore, the models of \refcite{DonevaEtAl2015} become unstable to collapse when the star radiates away up to 20\% of its angular momentum, so we considered rotation rates greater than $0.8\,\Omega_\mathrm{K}$.  A model without an enhanced Kepler limit is also shown, for comparison.

\begin{figure*}
	\includegraphics[width=\textwidth,keepaspectratio]{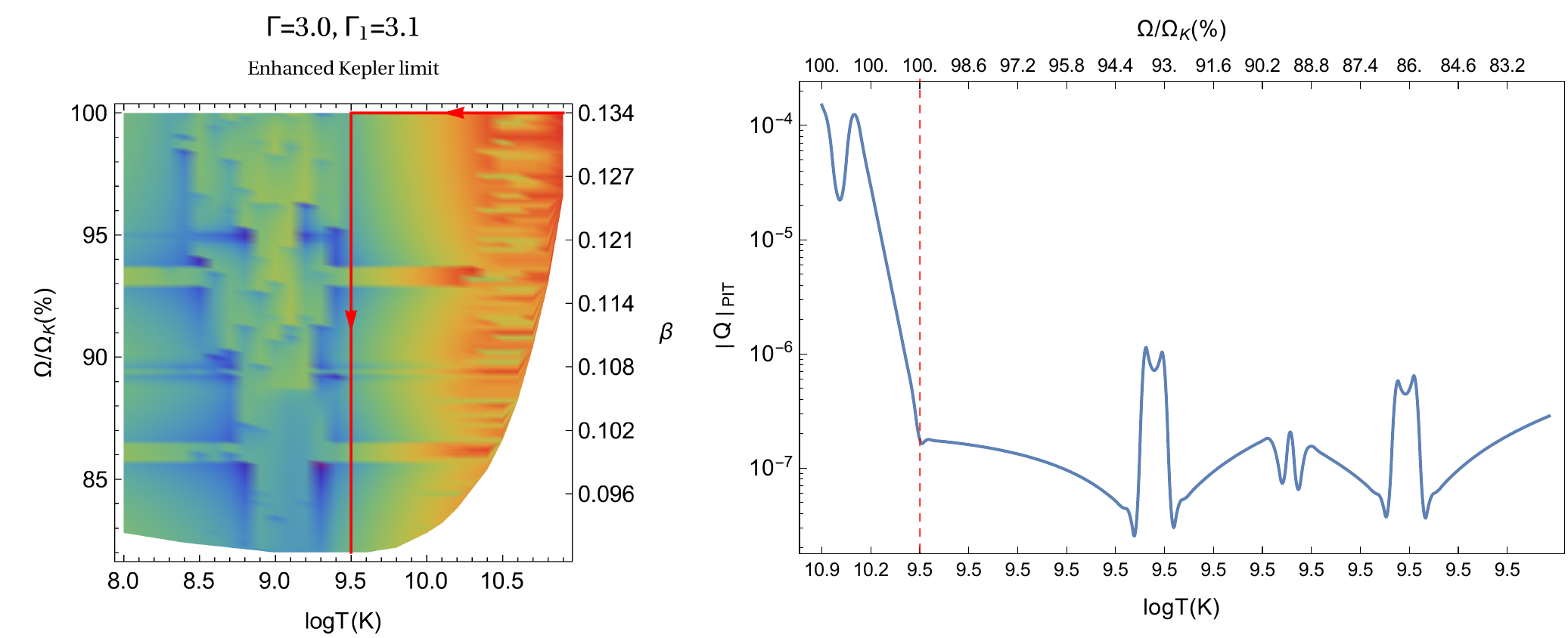}
	\caption{Hypothetical evolution of our toy model of a merger-derived, supramassive neutron star with $M=2.5\,M_\odot$ and $R=12\,\mathrm{km}$, through the instability window of the $l=m=2$ $f$-mode (\textit{left}), and the corresponding evolution of the lowest stable PIT (\textit{right}). The star obeys a polytropic equation of state with $\Gamma=3$ and an adiabatic exponent $\Gamma_1=3.1$. In this example, the star enters the window during its cooling phase, rotating at its maximum angular velocity, until thermal equilibrium is established (indicated by the vertical dashed line), at which point it descends the window at $T\approx 3\times10^9\,\mathrm{K}$ \cite{DonevaEtAl2015}.}
	\label{fig:2f_PIT_MDNS_hypothetical_evolution}
\end{figure*}

Finally, a hypothetical (based on the results of \refcite{DonevaEtAl2015}) evolution of a star inside the instability window of the quadrupole $f$-mode is shown in \cref{fig:2f_PIT_MDNS_hypothetical_evolution}, along with the variation of the mode's saturation amplitude.

%-----------------------------------------------------------------------------------------------------------------------------%

\subsubsection{Discussion}

\paragraph{Features.} The same features that were discussed in the previous section for supernova-derived stars can also be seen in \cref{fig:2f_PIT_MDNS,fig:3f_PIT_MDNS}. The same reasoning can be used to explain the characteristic decrease of the saturation amplitude from the edge to the interior of the instability window, as well as the horizontal bands that appear at certain angular velocities.

The fact that the windows of supramassive stars are larger, compared to their counterparts from supernova-derived stars, justifies the increase of the maximum value that the saturation amplitude can attain: according to \cref{f-g coupling saturation amplitude scaling}, the saturation amplitude scales with the daughter $g$-mode's damping rate, which can achieve greater (absolute) values at higher or lower temperatures.

An additional feature, observed only in the case of supramassive stars (and mainly in the results for the quadrupole $f$-mode in \cref{fig:2f_PIT_MDNS}), is this vertical brushstroke-like structure at intermediate temperatures. The anomalous behavior of the saturation amplitude in this area occurs, in a similar manner to the horizontal band feature, due to daughter pair changes: a daughter pair which can successfully saturate the parent fails to do so once the star enters this area. The reason is that some daughter $g$-modes become CFS-unstable inside this region and can no longer stop the parent's growth (remember that the daughter modes have to be stable). The unstable parent will then get saturated by a different daughter pair, which may lead to a sudden change of the saturation amplitude.

\begin{figure*}
	\begin{tabular}{ccc}
		(a) & (b) & (c) \\
		\includegraphics[width=.345\textwidth,keepaspectratio]{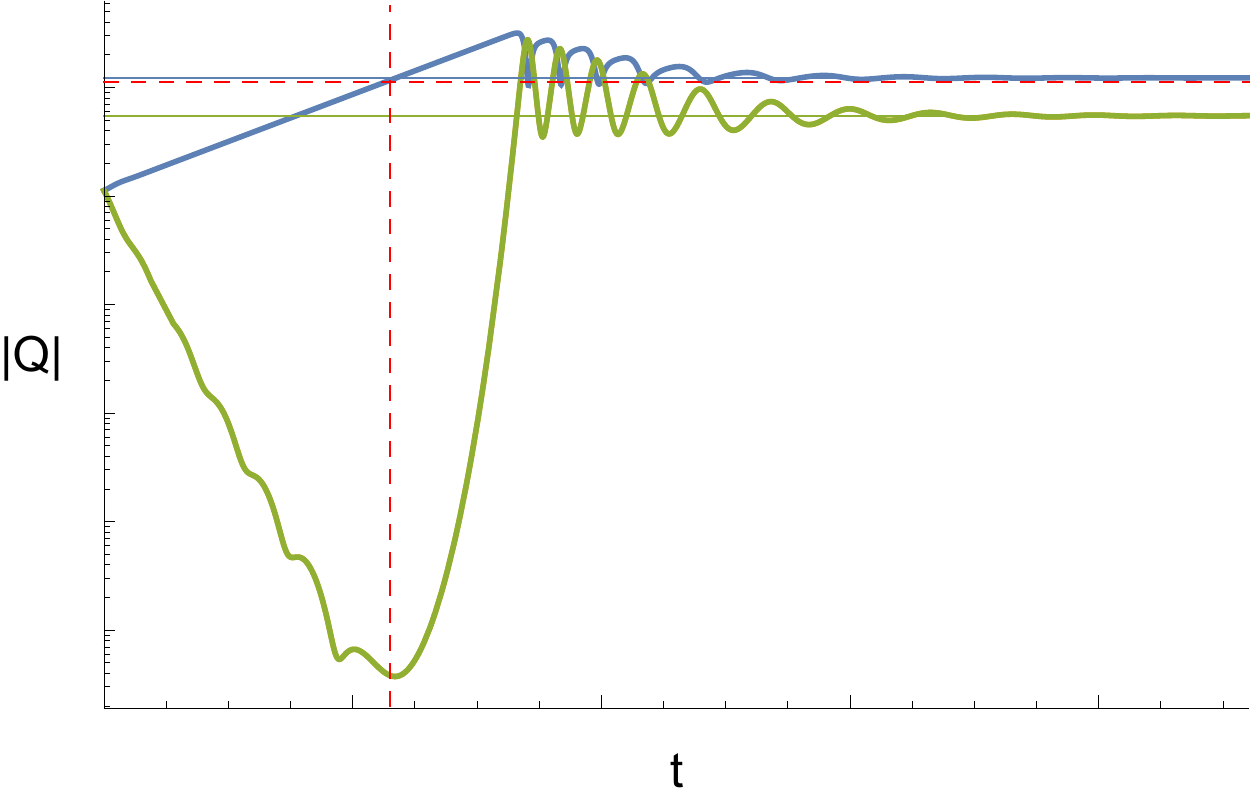} & \includegraphics[width=.3225\textwidth,keepaspectratio]{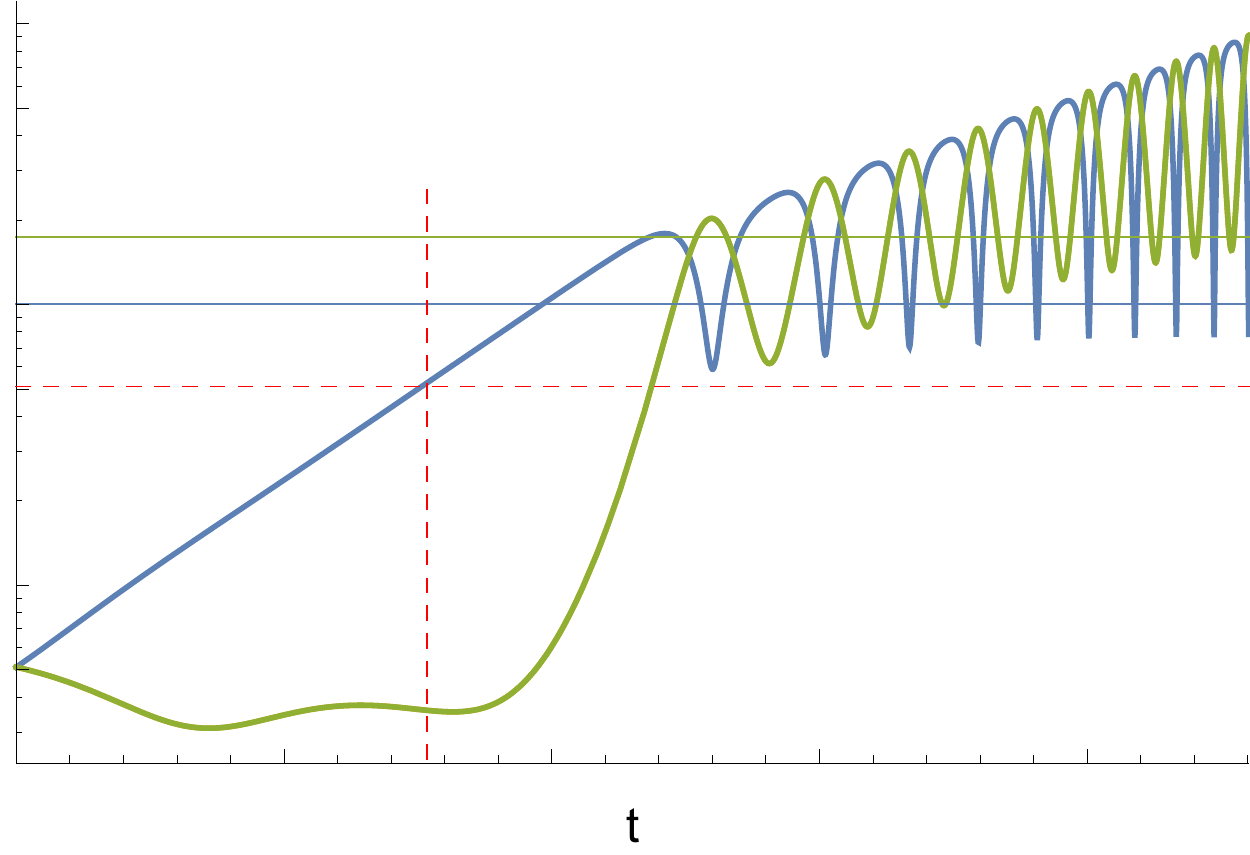} & \includegraphics[width=.3225\textwidth,keepaspectratio]{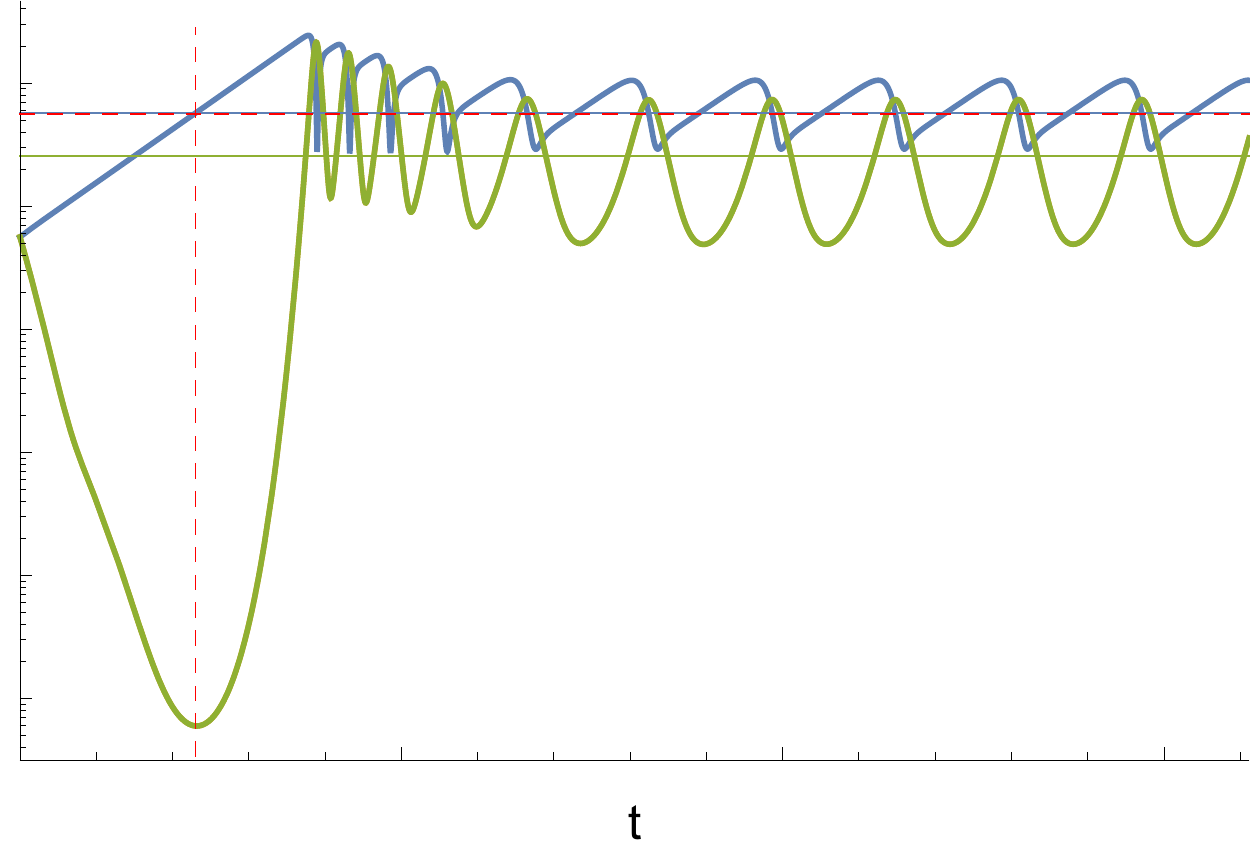}
	\end{tabular}
	\caption{Evolution of a parametrically resonant triplet with two identical daughters. Horizontal solid lines denote the equilibrium amplitudes of the modes, whereas the horizontal and vertical dashed lines indicate the PIT and the PIT-crossing time respectively. (a) Successful saturation, with both stability conditions \eqref{saturation condition 1} and \eqref{saturation condition 2} satisfied $(\gamma_\beta/\gamma_\alpha=\gamma_\gamma/\gamma_\alpha=-5,\,\Delta\omega/\gamma_\alpha=20)$. (b) Unsuccessful saturation, with condition \eqref{saturation condition 1} being false $(\gamma_\beta/\gamma_\alpha=\gamma_\gamma/\gamma_\alpha=-1/3,\,\Delta\omega/\gamma_\alpha=10/3)$. (c) Quasi-successful saturation, with condition \eqref{saturation condition 2} being false $(\gamma_\beta/\gamma_\alpha=\gamma_\gamma/\gamma_\alpha=-5,\,\Delta\omega/\gamma_\alpha=5)$.}
	\label{fig:parametrically resonant system}
\end{figure*}

\paragraph{Properties.} As previously mentioned, we enhanced the Kepler limit of our Newtonian polytropic models, in order to reproduce the growth time scales of the models used in \refcite{DonevaEtAl2015}. The factors leading to so short growth time scales in the latter are relativity\footnote{If the Cowling approximation, used by \refcite{DonevaEtAl2015}, were dropped, the instability should be amplified even more \cite{StavropoulosPrivCom2016}, because it sets in at smaller rotation rates \cite{ZinkEtAl2010}.} \cite{GaertigEtAl2011}, realistic equations of state \cite{DonevaEtAl2013}, and, of course, the large masses and angular momenta of these supramassive stars. The behavior of the angular momentum is an important subtlety of such stars. As shown by \refcite{CookEtAl1992}, there are regions where the loss of angular momentum spins the star \emph{up}.\footnote{The angular velocity increases, but the moment of inertia decreases more rapidly.} This feature cannot be mimicked by our Newtonian polytropes (see, e.g., \refcite{LaiEtAl1993}). As a result, we get the same $\tau_\mathrm{GW}$ with \refcite{DonevaEtAl2015} at the (enhanced) Kepler limit, but not for lower angular momenta, for which our growth time scale becomes significantly longer. This does not happen in the models of \refcite{DonevaEtAl2015} which, as described above, spin up instead when angular momentum is lost due to gravitational wave emission, inducing only a slight increase in $\tau_\mathrm{GW}$. This is why \refcite{DonevaEtAl2015} uses the ratio of kinetic to gravitational potential energy, $\beta$, instead of $\Omega$, to parametrize rotation. In \cref{fig:2f_PIT_MDNS,fig:3f_PIT_MDNS} we use both parameters: $\Omega$, as a reminder of the Newtonian origin of the calculation, and $\beta$, to indicate the connection with the realistic model.

The fact that the gravitational wave growth time scale of the parent mode should stay approximately the same throughout the window is not expected to change the results about the saturation amplitude. Although in our models $\tau_\mathrm{GW}$ changes by orders of magnitude for different angular momenta, it is not included in the evaluation of \cref{PIT}. The parent's growth rate $\gamma_\alpha$ affects the couplings indirectly, through the saturation conditions \eqref{saturation condition 1} and \eqref{saturation condition 2}. This means that even if $\tau_\mathrm{GW}$ had its Keplerian value everywhere, then, since the parent is successfully saturated at the Kepler limit by some daughter pairs, there will always be daughters with similar properties which will saturate it at lower angular momenta as well.

%-----------------------------------------------------------------------------------------------------------------------------%
%%%%%%%%%%%%%%%%%%%%%%%%%%%%%%%%%%%%%%%%%%%%%%%%%%%%%%%%%%%%%%%%%%%%%%%%%%%%%%%%%%%%%%%%%%%%%%%%%%%%%%%%%%%%%%%%%%%%%%%%%%%%%%%
%-----------------------------------------------------------------------------------------------------------------------------%

\section{The saturation mechanism} \label{sec:The saturation mechanism}

%-----------------------------------------------------------------------------------------------------------------------------%
%-----------------------------------------------------------------------------------------------------------------------------%

\subsection{Possible evolutions} \label{subsec:Possible evolutions}

As mentioned in \cref{subsec:Saturation}, the lowest PIT is, in most cases, stable, i.e., the triplet amplitudes converge around their equilibrium solution and saturation is considered successful (\cref{fig:parametrically resonant system}\hyperref[fig:parametrically resonant system]{a}). This is guaranteed by a condition involving the growth/damping rates and the detuning, given by Eq.~(4.24) in Paper I (see also Fig.~4 therein) and roughly approximated by \cref{saturation condition 1,saturation condition 2}. The impact of these constraints on the parametrically resonant system \eqref{equations of motion with normalisation choice} has been studied by various authors (e.g., \refcite{WersingerEtAl1980,*WersingerEtAl1980b,Dimant2000}), who discovered interesting behaviors throughout the parameter space.

The significance of \cref{saturation condition 1} is fairly easy to see: the daughters have to dissipate the incoming energy from the parent faster than the parent grows. Otherwise, the parent's amplitude keeps growing (at a rate lower than $\gamma_\alpha$), dragging the daughters along as it does so. The three modes diverge from their equilibria, by constantly exchanging energy at an increasing frequency. The energy of the system grows at a rate \cite{Dziembowski1982}
\begin{equation}
	\der{}{t}\sum_i |Q_i|^2 E_\mathrm{unit}=\sum_i 2\gamma_i |Q_i|^2 E_\mathrm{unit} \label{system energy rate}
\end{equation} 
$(i=\alpha,\beta,\gamma)$ and saturation fails (\cref{fig:parametrically resonant system}\hyperref[fig:parametrically resonant system]{b}).

On the other hand, \cref{saturation condition 2} poses a surprising constraint, by demanding that the detuning have a lower limit. When this condition is not satisfied [but \cref{saturation condition 1} is], a rich variety of evolutions can occur, depending on the values of the parameters. Growing solutions may still appear for small values of the damping rates or the detuning, but bounded evolutions dominate throughout the rest of the parameter space. \Refcite{WersingerEtAl1980,*WersingerEtAl1980b} report the appearance of limit cycles, with periods\footnote{A cycle with period $n$ intersects the Poincar\'e section $n$ times \cite{WersingerEtAl1980,*WersingerEtAl1980b}.} ranging from 1 to 32, as well as chaotic orbits, where the amplitudes of the modes oscillate around their equilibrium values (quasi-successful saturation).

The simplest case, of a limit cycle with period 1, was more thoroughly examined in \refcite{Moskalik1985} and is shown in \cref{fig:parametrically resonant system}\hyperref[fig:parametrically resonant system]{c}. The time scale of the modulation is $\sim\gamma_\alpha^{-1}$ and its peak-to-peak depth mainly depends on the ratio $|\Delta\omega/\gamma_\beta|$, for a triplet with two identical daughters $(\gamma_\beta=\gamma_\gamma)$. The modulation is larger when $|\Delta\omega/\gamma_\beta|$ is small, i.e., when i) the three modes are close to resonance $(\Delta\omega=\omega_\alpha-\omega_\beta-\omega_\gamma\approx 0)$, or ii) the daughters are strongly damped. The latter may sound unexpected, but, as seen from \cref{equations of motion with normalisation choice}, large daughter damping rates ``delay'' the nonlinear terms from becoming significant, thus allowing the parent to reach higher amplitude values. This characteristic probably explains why triplets with small detunings need to also have small daughter damping rates in order to 
successfully saturate (see Fig.~4 in Paper I).

Since the validity of \cref{saturation condition 2} is not, usually, necessary for the system to saturate, one may ask why did we always take it into account. With the exception of certain cases (e.g., $\gamma_\beta=\gamma_\gamma,\,|\gamma_\beta|\gg|\gamma_\gamma|$), the approximate relations \eqref{saturation condition 1} and \eqref{saturation condition 2} are hard to disentangle from the more general Eq.~(4.24) of Paper I. Whenever this condition is false, it is not easy to systematically check if this happens because of small daughter damping rates or a low detuning. This is the reason we considered the lowest \emph{stable} PIT as the saturation point.

Having calculated the lowest PITs throughout the instability window, the only secure way to determine whether they lead to bounded evolutions is to solve \cref{equations of motion with normalisation choice} for every one of them (up to $\approx 2600$ for some models). This is beyond the scope of our approach, but also quite unnecessary, because, in their vast majority, the lowest PITs are stable. As we shall see in \cref{sec:Comparison with r-modes}, this does not happen for the couplings of the unstable $r$-mode.

%-----------------------------------------------------------------------------------------------------------------------------%
%-----------------------------------------------------------------------------------------------------------------------------%

\subsection{Frequency synchronization} \label{subsec:Frequency synchronization}

\begin{figure}
	\includegraphics[width=.48\textwidth,keepaspectratio]{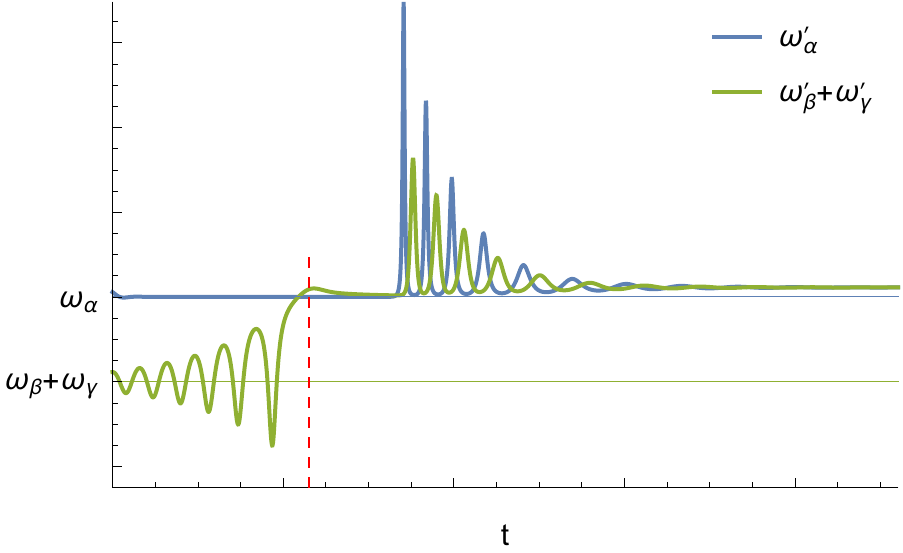}
	\caption{Evolution of the shifted (due to nonlinear coupling) mode frequencies towards nonlinear resonance, for the triplet of \cref{fig:parametrically resonant system}\hyperref[fig:parametrically resonant system]{a}. Horizontal lines denote the frequencies of the linear system and the vertical dashed line indicates the PIT-crossing time.}
	\label{fig:nonlinear_frequency_evolution}
\end{figure}

As discussed above, the detuning $\Delta\omega$ of the resonant triplet affects the evolution of the system in a notable manner. Interestingly enough though, this mismatch between the mode frequencies is compensated by nonlinear effects. Applying a procedure outlined in Sec.~IV\,F of Paper I (from which all the relations that follow can be derived), we split the (complex) amplitude $Q_i$ into its real amplitude and phase components, $|Q_i|$ and $\vartheta_i$. The harmonic time dependence of the mode becomes $\exp[i(\omega_i t+\vartheta_i)]$, resulting in a frequency shift
\begin{equation}
	\omega_i'=\omega_i+\dot{\vartheta}_i, \label{frequency shift}
\end{equation}
with
\begin{equation}
	\dot{\vartheta}_i=\frac{\omega_i \mathcal{H}}{E_\mathrm{unit}}\frac{|Q_\alpha Q_\beta Q_\gamma|}{|Q_i|^2}\cos\varphi, \label{mode i phase derivative}
\end{equation}
where $\varphi=\vartheta_\alpha-\vartheta_\beta-\vartheta_\gamma+\Delta\omega t$. Then, solving for the equilibrium of \cref{equations of motion with normalisation choice}, we get $\dot{\varphi}=0$, or
\begin{equation}
	\omega_\alpha'=\omega_\beta'+\omega_\gamma'. \label{nonlinear resonance}
\end{equation} 
This has been referred to as frequency synchronization \cite{Aikawa1984} or phase lock \cite{DziembowskiKovacs1984}, and is an anticipated effect of nonlinear resonance. Replacing equilibrium values in \cref{mode i phase derivative} we find that the frequency shift is
\begin{equation}
	\omega_i'=\omega_i-\left|\gamma_i\right| \frac{\Delta\omega}{\gamma_\alpha+\gamma_\beta+\gamma_\gamma}. \label{frequency shift at equilibrium}
\end{equation}
It should be noted that \cref{nonlinear resonance,frequency shift at equilibrium} are valid only in the case of successful saturation, when the mode amplitudes are constants.\footnote{In principle, they could also apply to the quasi-stable equilibria discussed above, as average-value relations \cite{Moskalik1985}.} The evolution of the shifted mode frequencies towards nonlinear resonance is shown in \cref{fig:nonlinear_frequency_evolution}, for the triplet of \cref{fig:parametrically resonant system}\hyperref[fig:parametrically resonant system]{a}.

In general, the frequency shift is negligible for the parent mode. Among lowest-PIT triplets, we found a maximum shift of $\sim 0.1\,\mathrm{Hz}$ for supernova-derived stars. For merger-derived stars, some couplings induce a frequency shift as high as $\sim 0.1\,\mathrm{kHz}$. Given that the $l=m$ $f$-modes have frequencies of a few $\mathrm{kHz}$, the latter could be significant. However, these couplings are very few, since one needs the parameters in \cref{frequency shift at equilibrium} finely tuned (large $\gamma_\alpha$ and $|\Delta\omega|$; $|\gamma_{\beta,\gamma}|$ as small as possible) to produce a considerable frequency shift. For the daughter modes, the frequency shift is larger, since they also have to cover the frequency mismatch $\Delta\omega$ (at most $\sim 0.1\,\mathrm{kHz}$) to catch up with the parent (see \cref{fig:nonlinear_frequency_evolution}).\footnote{From \cref{saturation condition 1,frequency shift at equilibrium} we see that the frequency shift always has the same sign with the detuning 
$\Delta\omega$, meaning that the parent frequency is always shifted away from the daughter frequencies.}

%-----------------------------------------------------------------------------------------------------------------------------%
%%%%%%%%%%%%%%%%%%%%%%%%%%%%%%%%%%%%%%%%%%%%%%%%%%%%%%%%%%%%%%%%%%%%%%%%%%%%%%%%%%%%%%%%%%%%%%%%%%%%%%%%%%%%%%%%%%%%%%%%%%%%%%%
%-----------------------------------------------------------------------------------------------------------------------------%

\section{Comparison with \texorpdfstring{$\bm{r}$}{r}-modes} \label{sec:Comparison with r-modes}

Soon after its discovery \cite{Andersson1998,FriedmanMorsink1998}, the $r$-mode instability was proposed as an explanation for the observed neutron star spin periods. As mentioned in \cref{subsec:Supernova-derived neutron stars}, angular velocities close to the Kepler limit should, in principle, be possible in nascent neutron stars, yet observations do not seem to confirm this. Furthermore, spin-up due to accretion in low-mass x-ray binaries (LMXBs) should also allow for rotation rates larger than the observed ones. In both cases, gravitational wave emission from unstable $r$-modes could prevent neutron stars from reaching millisecond periods \cite{LindblomEtAl1998,AnderssonEtAl1999,AnderssonEtAl1999b,AnderssonEtAl2000}.

Apart from determining the strength (and hence the detectability) of the generated gravitational wave signal, the saturation amplitude of the unstable $r$-mode is important for the evolution of the neutron star inside the instability window (a process which sets its final spin rate), because it affects the spin-down time scale. The saturation energy of the $r$-mode was initially taken to be of order the rotational energy of the star \cite{OwenEtAl1998} $(E_\mathrm{unit}=MR^2\Omega^2/2)$. Later studies, which will be reviewed below, showed that nonlinear coupling saturates the $r$-mode at much lower amplitudes.

After Schenk \textit{et al.} \cite{SchenkEtAl2001} laid the groundwork, by deriving a consistent mode coupling formalism, Morsink \cite{Morsink2002} calculated nonlinear couplings between $r$-modes in a stratified polytrope. Then, Arras \textit{et al.} \cite{ArrasEtAl2003} provided the first analytic estimate of the saturation amplitude, by coupling the $l=m=2$ $r$-mode to other inertial modes, in a star with negligible buoyancy. After calculating the modes in the WKB limit, they considered two cases: i) the discrete (or ``weak-driving'') limit and ii) the continuum (or ``strong-driving'') limit. 

The first case is identical to the one we study here: the unstable mode grows and, after crossing the lowest PIT, excites the corresponding daughter pair and saturates. Thus, the number of modes involved is small and the mode spectrum can be regarded as discrete. For saturation to occur, the saturation conditions \eqref{saturation condition 1} and \eqref{saturation condition 2} have to be satisfied for the coupled triplet.\footnote{Arras \textit{et al.} are relaxing the saturation conditions by allowing the second one [\cref{saturation condition 2}] not to be true. The reason behind this was explained in \cref{sec:The saturation mechanism}. \label{foot:Arras et al. saturation conditions}} In this sense, according to \cref{saturation condition 1}, the parent is weakly driven, compared to the daughters' damping rates. Although their estimations about the saturation amplitude are really low (see Sec.~6 of their paper), Arras \textit{et al.} conclude that the discrete limit is not a good approximation, neither 
for nascent nor for accreting neutron stars, because a large number of daughter modes is expected to be excited. This brings us to the continuum limit.

In the second case, the modes are treated as a continuum, since a large number of daughter modes is excited. This happens when the coupled triplets fail to satisfy the saturation condition \eqref{saturation condition 1}, so the driving rate of the parent is larger than the damping rates of the daughters. As the parent grows, it crosses many PITs but cannot reach a stable equilibrium. However, as it couples with more and more daughters, a turbulent energy cascade is formed and its growth rate decreases, until it finally settles and saturates. Arras \textit{et al.} find that the saturation energy is given by $E_\mathrm{mode}/E_\mathrm{unit}=10^{-6}(\alpha_e/0.1)\nu^5_\mathrm{kHz}$, where $E_\mathrm{unit}=MR^2\Omega^2/2$, $\nu$ is the spin rate of the star (measured in $\mathrm{kHz}$) and $\alpha_e$ parametrizes some ``uncertainties'' of their approach. Although taken to be $\approx 0.1$, $\alpha_e$ could be as low as $\sim 10^{-4}$ (see Sec.~2 of their paper), which would make the saturation energy even lower. 
Since we take $E_\mathrm{unit}=Mc^2$ throughout this paper, we can use Eq.~(4.13) from Paper I and find that the saturation energy above is approximately two orders of magnitude lower in our units.

Our study shows that the $f$-mode falls into the discrete limit case: the lowest PIT is usually stable and successfully saturates the mode. Comparing our \cref{f-g coupling saturation amplitude scaling} to Arras \textit{et al.}'s Eqs.~(78) and (80) we see that the temperature scalings do not change: if bulk viscosity dominates the damping, their saturation amplitude scales as $T^6$, whereas, if shear viscosity dominates, it scales as $T^{-2}$ (their $A$ is equivalent to our $|Q|$). Although our scalings are different by a factor of $1/2$ in power, this should not be a surprise. As explained in \cref{subsec:Supernova-derived neutron stars}, the temperature dependence in \cref{f-g coupling saturation amplitude scaling} is due to the $g$-mode daughter damping rate changing with temperature (as opposed to the $f$-mode daughter damping rate). On the other hand, Arras \textit{et al.} extract their Eqs.~(78) and (80) assuming two identical daughters, in which case the PIT is approximated by \cref{PIT gamma_beta=gamma_gamma}, where the daughter damping rate is squared.\footnote{They also ignore the term $\Delta\omega/(\gamma_\beta+\gamma_\gamma)$, see \cref{foot:Arras et al. saturation conditions}.}

The analytical work of Arras \textit{et al.} was followed by the simulations of Brink \textit{et al.} \cite{BrinkEtAl2004,*BrinkEtAl2004b,*BrinkEtAl2005}, where the saturation amplitude of the unstable $r$-mode was found ``experimentally''. Brink \textit{et al.} used an incompressible, homogeneous star (Maclaurin spheroid), which permits the analytical calculation of eigenfrequencies, eigenfunctions, and all the quantities that involve them (growth/damping rates, coupling coefficients). As opposed to Arras \textit{et al.}, they dropped the Cowling approximation, which, as they concluded, would otherwise neglect important terms in the couplings coefficients.

Their simulations included inertial modes with $l\le 29$ ($\approx 5000$ modes), which resulted in $\approx 1.5\times 10^5$ direct couplings to the $l=m=2$ $r$-mode, plus a large number of couplings among daughters themselves (daughter-daughter couplings). Starting with the integration of the triplet with the lowest PIT, via \cref{equations of motion with normalisation choice}, they gradually added more couplings, as the $r$-mode kept growing and more modes were rising above the noise level. With this technique, i.e., following the evolution of the mode amplitudes, one can achieve much longer integration times than ordinary hydrodynamical simulations, where the integration time step is set by the oscillation periods of the modes. 

In their work, they studied three types of large (i.e., involving all modes) systems: the conservative (Hamiltonian) system, as well as the strongly- and weakly-damped nonconservative system.

In the conservative system, the growth/damping rates are zero and, hence, the modes simply interact nonlinearly. This is known as the Fermi-Pasta-Ulam (FPU) problem and its extensive study has shown that, after the initial excitation of a large scale mode, a state of energy equipartitioning is reached, should the initial amplitude be larger than some threshold. This was indeed observed by Brink \textit{et al.}, with the equipartition time scale decreasing as the initial $r$-mode amplitude was increased.

The strongly-damped system corresponds to Arras \textit{et al.}'s weak-driving limit. Brink \textit{et al.} showed that, in this case, the mode amplitude evolution resembles that of a single triplet: the daughters' damping rates are large enough to halt the parent's growth, after the lowest PIT is crossed.\footnote{Like Arras \textit{et al.}, Brink \textit{et al.} also ignore the saturation condition \eqref{saturation condition 2} (see \cref{foot:Arras et al. saturation conditions}). This condition is not satisfied by the triplets with the lowest PITs in their simulations and, as a result, saturation occurs in the form of limit cycles or aperiodic motions (see \cref{sec:The saturation mechanism}).}

In the weakly-damped system, which in turn corresponds to the strong-driving limit of Arras \textit{et al.}, the situation is significantly more complex. The daughters are now not damped enough to stop the growth of the $r$-mode. However, daughter-daughter couplings distribute the incoming energy to many modes, thus preventing the $r$-mode from growing far beyond the second lowest PIT (Figs. 12 and 13 in their last paper). In fact, at this amplitude the rate to equipartition is similar to the $r$-mode's growth rate, making the contribution of the FPU mechanism to the damping of the instability quite important.

The simulations above were performed for a star rotating at $\approx 0.6\,\Omega_\mathrm{K}$ (the $r$-mode instability window minimum lies at a few percent of the mass-shedding limit \cite{LindblomEtAl1998}). The strongly-damped system resides at low temperatures $(T\sim 10^6-10^7\,\mathrm{K})$, whereas the weakly-damped system at intermediate temperatures $(T\sim 10^8-10^9\,\mathrm{K})$. At these temperatures, shear viscosity is the dominant damping mechanism (bulk viscosity is zero for an incompressible star).

Although Arras \textit{et al.}'s cascade picture was mostly confirmed in the weakly-damped regime, the saturation amplitudes reported by Brink \textit{et al.} are lower. The $r$-mode saturation energy was found to be $E_\mathrm{mode}/E_\mathrm{unit}\approx 10^{-10}-10^{-8}$, where $E_\mathrm{unit}=MR^2\Omega^2/2$, with the higher values occurring at the weakly-damped system and the low-temperature end of the strongly-damped system. Since our system is strongly-damped at all temperatures, this is in agreement with our results, where the saturation amplitude obtains larger values at low and high temperatures.

Based on the work of Brink \textit{et al.}, a series of simulations were performed by Bondarescu \textit{et al.} \cite{BondarescuEtAl2007,*BondarescuEtAl2009}, where the evolution of the neutron star through the $l=m=2$ $r$-mode instability window was studied. Since the $r$-mode instability is relevant both for accreting and nascent neutron stars, both cases were examined.

Bondarescu \textit{et al.} used a $\Gamma=2$ polytrope, assuming the couplings of Brink \textit{et al.}'s incompressible model. In particular, they incorporated the coupling with the lowest PIT (as obtained from Brink \textit{et al.}'s simulations) into the spin and temperature evolution equations. This makes the neutron star evolution more ``dynamical'', as opposed to previous work, where the $r$-mode saturation amplitude was treated as an arbitrary constant. In most cases, the lowest-PIT coupling was enough to stop the $r$-mode from growing, which places the star in Brink \textit{et al.}'s strongly-damped regime. It should be noted that, in addition to shear viscosity, Bondarescu \textit{et al.} also considered viscosity at the crust-core boundary layer and hyperon bulk viscosity, thus ``enhancing'' the dissipation effects.

By varying certain properties of the star, like the hyperon superfluid transition temperature, the fraction of the star above the threshold for direct Urca reactions, and the crust-core slippage factor, a number of interesting scenarios occur. Essentially, these quantities parametrize the strength of viscous and cooling effects. Except for their runaway evolutions, where the $r$-mode grows beyond the lowest PIT and the three-mode system fails, saturation always occurs at the lowest PIT. The latter slowly changes during the evolution, due to the temperature dependence of the daughters' damping rates. For the rest of the quantities comprising the PIT [\cref{PIT}], Bondarescu \textit{et al.} assume the values of the lowest-PIT triplet as ``statistically relevant'' constants.

Taking a step further, we calculated the saturation amplitude of the unstable $f$-modes, due to the three-mode coupling mechanism, \emph{throughout} their instability windows. By doing this, we extract the whole coupling spectrum of the $f$-modes, which enables us follow the different kinds of couplings and their strengths. Doing this for the $r$-mode will, most probably, not change the main results dramatically, because inertial modes are confined in a relatively small frequency range $[-2\Omega,2\Omega]$ and fine resonances are fairly easy to obtain. On the other hand, in the case of the $f$-mode there are different kinds of daughter pairs ($f$-$g$ or $g$-$g$, see \cref{subsec:Supernova-derived neutron stars}), which makes it more subtle to be described by use of ``effective'' values for parameters like the detuning or the coupling coefficient.

% The evolutions of Bondarescu \textit{et al.} showed that a gravitational wave signal from an unstable $r$-mode could be detectable with Advanced LIGO from sources within the local galactic group $(\sim 1\,\mathrm{Mpc})$.

%-----------------------------------------------------------------------------------------------------------------------------%
%%%%%%%%%%%%%%%%%%%%%%%%%%%%%%%%%%%%%%%%%%%%%%%%%%%%%%%%%%%%%%%%%%%%%%%%%%%%%%%%%%%%%%%%%%%%%%%%%%%%%%%%%%%%%%%%%%%%%%%%%%%%%%%
%-----------------------------------------------------------------------------------------------------------------------------%

\section{Summary} \label{sec:Summary}

We have presented the first results about the saturation of the $f$-mode instability in neutron stars, due to quadratic mode coupling. Using Newtonian polytropes to describe both supernova- and merger-derived neutron stars, we calculated all the couplings of the most unstable $f$-mode multipoles to other polar modes and obtained their saturation amplitude throughout their instability windows.

Once the fast-rotating, nascent neutron star enters the instability window, the $f$-mode is driven unstable due to the emission of gravitational waves (CFS instability). Coupling of the exponentially growing (parent) mode to damped (daughter) modes leads to a parametric resonance instability, during which a pair of daughters drain the parent's energy and grow, when the latter crosses their characteristic parametric instability threshold. If the daughters are sufficiently dissipated, the triplet reaches an equilibrium and saturates.

The efficiency of the coupling among three modes is determined by their coupling coefficient, depending on their eigenfunctions, and by their detuning, which measures how close to resonance they are. The triplet amplitudes approach constant values only if the detuning is larger than some lower limit, thus favoring far-from-exact resonances. As a result of the coupling, however, the mode frequencies are shifted and finally evolve towards an exact nonlinear resonance. On the other hand, a lower detuning usually induces oscillations in the triplet amplitudes, around their equilibrium values, in the form of limit cycles or chaotic orbits.

Although it is usually treated as a constant, the saturation amplitude changes throughout the window, due to its temperature dependence and because different daughter pairs may set the lowest PIT at different points. We found that the saturation amplitude is larger near the low- and high-temperature edges of the instability window (as high as $\approx 3\times 10^{-4}$), and gradually decreases at intermediate temperatures (with values as low as $\sim 10^{-9}$; the definition used for the amplitude is $|Q|=\sqrt{E_\mathrm{mode}/Mc^2}$\,).

The perturbative nonlinear approach that we use is, in its core, simple and has many advantages. As long as the eigenfrequencies and eigenfunctions of the modes are provided, it allows us to easily identify the important couplings in the system and precisely track their effects on the modes' amplitude evolution. Furthermore, it helps us reveal and understand the richness of possible outcomes and offers a strong insight into the problem, letting us follow every parameter's contribution.

The calculation of the eigenfrequencies and eigenfunctions of the modes, however, can be a quite laborious task, with analytic solutions existing only in simple models (e.g., homogeneous star) and with no natural limit on the number of modes that should be considered---for instance, solar observations have shown very high $p$-mode multipole oscillations. In order to obtain as many modes as possible, the slow-rotation approximation was utilized, which is the origin of the major uncertainties in our results (correctness of models aside).

The Newtonian formalism provides an accurate qualitative description of the problem, at least for supernova-derived neutron stars. In principle, general relativity should change some key components of the setup (e.g., larger instability windows, shorter growth time scales for the parent), thus affecting the final results. Moreover, it is the only appropriate framework for modeling supramassive post-merger remnants, since our Newtonian calculation reflects only their rudimentary properties. Therefore, a relativistic quadratic perturbation scheme needs to be developed in order to obtain conclusive results, especially considering that relativistic hydrodynamic simulations are still far from remaining stable during the secular time scales needed for the instability to grow.

Gravitational waves from neutron star oscillations will shed some light on the equation of state of dense matter. Signals generated by the $f$-mode instability might be detectable even with Advanced LIGO, from sources in the Virgo cluster, considering the highest value of the saturation amplitude obtained here \cite{PassamontiEtAl2013,DonevaEtAl2015}. The gravitational-wave era in astronomy has only begun and much work still has to be done, regarding the elimination of the major uncertainties and the improvement of the models, in order to reach confident conclusions.

%--------------------Acknowledgments----------------------%

%\begin{acknowledgments}
	
%\end{acknowledgments}

%-----------------------Appendix--------------------------%

\revappendix*

\let\oldsection=\section
\def\section#1{\oldsection{\uppercase{#1}}}               % Capitalise appendix section title

%-----------------------------------------------------------------------------------------------------------------------------%
%%%%%%%%%%%%%%%%%%%%%%%%%%%%%%%%%%%%%%%%%%%%%%%%%%%%%%%%%%%%%%%%%%%%%%%%%%%%%%%%%%%%%%%%%%%%%%%%%%%%%%%%%%%%%%%%%%%%%%%%%%%%%%%
%-----------------------------------------------------------------------------------------------------------------------------%

\section{Approximate relations for the parametric instability threshold} \label[appsec]{sec:Approximate relations for the parametric instability threshold}

We are going to examine \cref{PIT} for two limiting cases: 1) one daughter mode is damped much more quickly than the other, and 2) the daughter modes are equally damped. For each case, we will further consider two additional limits: a) the detuning equals the daughters' damping rates, and b) the detuning is much larger than the daughters' damping rates. The limit in which the detuning is much smaller than the daughters' damping rates is inconsistent with the saturation condition \eqref{saturation condition 2}.\footnote{Based on the discussion in \cref{sec:The saturation mechanism} one could also consider this case, but it differs at most by a factor of 2 with case (a).}

%-----------------------------------------------------------------------------------------------------------------------------%
%-----------------------------------------------------------------------------------------------------------------------------%

\subsection{\texorpdfstring{$\bm{|\gamma_\beta|\gg|\gamma_\gamma|}$}{Different daughter damping rates}} \label[appsec]{subsec:Different daughter damping rates}

If one daughter's damping rate is much larger, \cref{PIT} becomes
\begin{equation}
	|Q_\mathrm{PIT}|^2\approx\frac{\gamma_\beta\gamma_\gamma}{\omega_\beta\omega_\gamma}\frac{E_\mathrm{unit}^2}{\mathcal{H}^2}\left[1+\left(\frac{\Delta\omega}{\gamma_\beta}\right)^2\right]. \label{PIT gamma_beta>>gamma_gamma}
\end{equation} 
We then take the two subcases:
\begin{enumerate}
	\item[1a.] $|\Delta\omega|\approx|\gamma_\beta|$
		\begin{equation}
			|Q_\mathrm{PIT}|^2\approx2\frac{\gamma_\beta\gamma_\gamma}{\omega_\beta\omega_\gamma}\frac{E_\mathrm{unit}^2}{\mathcal{H}^2}, \label{PIT gamma_beta>>gamma_gamma Deltaomega=gamma_beta}
		\end{equation}
	\item[1b.] $|\Delta\omega|\gg|\gamma_\beta|$
		\begin{equation}
			|Q_\mathrm{PIT}|^2\approx\frac{\gamma_\gamma}{\gamma_\beta}\frac{\Delta\omega^2}{\omega_\beta\omega_\gamma}\frac{E_\mathrm{unit}^2}{\mathcal{H}^2}. \label{PIT gamma_beta>>gamma_gamma Deltaomega>>gamma_beta}
		\end{equation} 
\end{enumerate}
The case $|\Delta\omega|\approx|\gamma_\gamma|$ is skipped, because this would mean $|\Delta\omega|\ll|\gamma_\beta|$.

%-----------------------------------------------------------------------------------------------------------------------------%
%-----------------------------------------------------------------------------------------------------------------------------%

\subsection{\texorpdfstring{$\bm{\gamma_\beta\approx\gamma_\gamma}$}{Same daughter damping rates}} \label[appsec]{subsec:Same daughter damping rates}

In cases when the daughter damping rates are the same, \cref{PIT} becomes
\begin{equation}
	|Q_\mathrm{PIT}|^2\approx\frac{\gamma_\beta^2}{\omega_\beta\omega_\gamma}\frac{E_\mathrm{unit}^2}{\mathcal{H}^2}\left[1+\left(\frac{\Delta\omega}{2\gamma_\beta}\right)^2\right]. \label{PIT gamma_beta=gamma_gamma}
\end{equation}
The two subcases additionally give:
\begin{enumerate}
	\item[2a.] $|\Delta\omega|\approx|\gamma_\beta|$
		\begin{equation}
			|Q_\mathrm{PIT}|^2\approx\frac{5}{4}\frac{\gamma_\beta^2}{\omega_\beta\omega_\gamma}\frac{E_\mathrm{unit}^2}{\mathcal{H}^2}, \label{PIT gamma_beta=gamma_gamma Deltaomega=gamma_beta}
		\end{equation}
	\item[2b.] $|\Delta\omega|\gg|\gamma_\beta|$
		\begin{equation}
			|Q_\mathrm{PIT}|^2\approx\frac{\Delta\omega^2}{4\,\omega_\beta\omega_\gamma}\frac{E_\mathrm{unit}^2}{\mathcal{H}^2}. \label{PIT gamma_beta=gamma_gamma Deltaomega>>gamma_beta}
		\end{equation} 
\end{enumerate}

%--------------------Bibliography-------------------------%
%
\bibliography{references}
%
%------------------------End------------------------------%
\end{document}